\documentclass[prb]{revtex4}
\usepackage{amsmath,amsfonts}

\begin{document}

\title{Decoherence of Schr\H{o}dinger cat states in a
  Luttinger liquid}

\author{P.~Degiovanni}
\email[Email: ]{Pascal.Degiovanni@ens-lyon.fr}
\author{S.~Peysson}
\email[Email: ]{Stephane.Peysson@ens-lyon.fr}
\address{Laboratoire de Physique de l'Ecole Normale
  Sup\'erieure de Lyon, Unit\'e de mixte de recherche CNRS et Ecole Normale
Sup\'erieure de Lyon (UMR~5672), Groupe de
  Physique Th\'eorique, 46 all\'ee d'Italie, 69364 Lyon
  Cedex 07, France.}

\pacs{71.10.Pm, 42.50.Md, 71.38.+i}

\begin{abstract}
Schr\H{o}dinger cat states built from
quantum superpositions of left or right Luttinger
fermions located at different positions in a spinless Luttinger
liquid are considered. 
Their decoherence rates are computed within the
bosonization approach using as environments the
quantum electromagnetic field or two or three dimensional
acoustic phonon baths. Emphasis is put on the differences
between the electromagnetic and acoustic environments.

LPENSL-Th-01/2000
\end{abstract}


\medskip

\maketitle

\section{Introduction and motivations}
\label{secIntro}

Decoherence is a very important issue for mesoscopic
systems since it governs the crossover between quantum
and quasi-classical transport regimes. 
Coherence of mesoscopic conductors gives
rise to various Aharonov-Bohm interference effects
such as permanent currents in a mesoscopic ring,
and conductance oscillations as a
function of the external magnetic field.

\medskip

The problem of electron decoherence in metals at zero
temperature is an active area of research both
from the theoretical and experimental point of
view. Recent experiments claim to observe a saturation of
the dephasing time $\tau_{\phi}$ at very low
temperatures \cite{Mohanty:1997-1}. 
Strong discussions among theorists arose
from these
observations\cite{Aleiner:1998-1,Golubev:1998-4,Aleiner:1999-1,Golubev:1999-1,Golubev:1999-2}. 
The heart of the debate,
summarized for example in Mohanty's recent letter \cite{Mohanty:1999-1}, is to
determine whether the conventional theory of dephasing in
Fermi liquids \cite{Altshuler:1982-1,Altshuler:1985-1,Chakravarty:1986-1} could explain the saturation of
$\tau_{\phi}$ (for example as an effect of an external 
microwave radiation \cite{Aleiner:1998-1,Cohen:1999-1}) or
whereas one should reconsider the 
theory completely \cite{Golubev:1998-1,Golubev:1998-2,Golubev:1999-3}. 
Although such an excitement is a strong motivation for
working on decoherence in disordered mesoscopic conductors, 
we shall present a model for studying {\em the decay of
Schr\H{o}dinger cat states in 1D ballistic conductors}.

\medskip

From a methodological point of view, our line of thought
is very close to the one used in atomic physics, for
example in theoretical works 
\cite{Davidovich:1996-1,Raimond:1997-1} on decoherence experiments in
cavity QED \cite{Brune:1996-1}. It relies heavily on the use of
simple exactly solvable models of decoherence, such as
the Caldeira-Leggett model \cite{CL:1983-2}. 
We think that this point of view brings a different light on the
question of decoherence by taking into account the
electron fluid as a whole and studying the coupling of
this many-body system to the external reservoirs. It
makes a bridge between atomic physics situations, where
decoherence is extremely well controlled and for
which simplified decoherence models apply almost
directly and mesoscopic conductors which are complex
interacting systems.
In particular, this point of view is well suited to the 1D case 
because of interactions. In a non-Fermi
liquid, one cannot keep track of an individual electron
because of orthogonality catastrophe effects. Moreover,
electromagnetic or acoustic radiation emitted by
one part of the system and absorbed by another part may
have drastic effect on the strongly correlated electron
state. These remarks motivated our method, based
on the study of the coupling of an external environment
to the low energy excitations of the 1D electron system.
Following general ideas of Stern, Aharonov and Imry~\cite{Stern:1990-1}, 
we have computed the decoherence rate of
Schr\H{o}dinger cat states built from localized
excitations ({\it e.g} Luttinger fermions) introduced at
different places in the system or at the same place but
moving in different directions (left and right moving components). 
As expected, we have found that such linear
superpositions decay into statistical mixtures even at zero
temperatures and computed various decoherence scenarii.
This is the main result of this paper and it means that {\em a
1D pure ballistic conductor exhibits decoherence at absolute
zero} in the sense of Schr\H{o}dinger cat states
decay. 

\medskip

One dimensional conductors in the ballistic regime are
appropriately described by an effective interacting
theory: the Luttinger liquid \cite{Haldane:1981-1}. In this effective theory,
interactions between electrons are put by
means of an electrostatic short-ranged potential.
Obviously, within the approximation of a
linear dispersion relation for particle-hole excitations,
the Luttinger effective theory contains no source for
decoherence. A coupling to an external quantum environment is necessary
to introduce decoherence in the Luttinger liquid. The quantized
electromagnetic field and a 2D or 3D bath of longitudinal
phonons will be considered in the
present paper. The combined use of bosonization and
non-equilibrium techniques make it possible 
to solve this problem. Within the bosonization framework,
the pioneering work \cite{Loss:1993-1}
by Martin and Loss investigates the question of
equilibrium permanent
currents induced by fluctuations of the quantized
electromagnetic field. They have shown that coupling the Luttinger liquid
to QED leads to a renormalization of the Luttinger liquid
parameters. Our discussion of {\em dynamical} and
therefore {\em non-equilibrium} aspects of the 
coupled Luttinger \& QED system will show how this
renormalization appears dynamically.
Let us mention that coupling a Luttinger liquid to
one dimensional phonons also renormalizes the Luttinger
liquid parameters and drives a 1D Fermi liquid to a
non Fermi liquid fixed point\cite{Loss:1994-1,Martin:1995-1,Martin:1995-2,Eggert:1996-1}. But a
1D phonon bath does not introduce any intrinsic
decoherence in the Luttinger liquid since, roughly
speaking, it does not have enough modes.
That's why 2D and 3D phonon baths are considered in this
paper and
we have shown that these baths have enough modes to kill
Schr\H{o}dinger cat states. 

\medskip

To be more precise on the results presented in this
paper, we have shown that the electromagnetic decoherence time
is much larger, although not infinite,
than the natural time associated with the Luttinger
system. This is mainly due to the weakness and the transversality of the
coupling between photons and the Luttinger liquid. 
Therefore the coherent Luttinger liquid paradigm is
not ruled out by its coupling to QED. We have also shown that the acoustic
decoherence is much stronger than the electromagnetic
one. This difference comes from the fact that many
bosonic modes of the Luttinger liquid
have an efficient acoustic radiation rate whereas only
very few modes dominate the decoherence process in the
electromagnetic case. In the acoustic case, decoherence
takes place over a much shorter time scale than
dissipation contrarily to the electromagnetic case.

\medskip

This paper is organized as follows: the model is presented and the
Feynman-Vernon and Keldysh basic tools are briefly
recalled in section
\ref{sectionQED}. Section \ref{secQBM} makes contact with
the Quantum Brownian Motion problem (QBM). The central problem of mutual
decoherence of Schr\H{o}dinger cat states is addressed in
section \ref{secDecLL}. Results are summarized and
possible extensions are  discussed in the conclusion.
All along the paper, the electromagnetic and acoustic
cases are discussed separately. Technical details are gathered in
appendices.

\section{Electron systems coupled to external reservoirs
  (QED or phonons)}
\label{sectionQED}

\subsection{The Feynman-Vernon-Keldysh method}

Studying an out of equilibrium quantum system boils down to the
computation of its density matrix as a function of time. A
functional integral approach to this problem has been
given long time ago in the context of perturbative field
theory by Keldysh \cite{Keldysh:1965-1}. 
Let $\rho (t) $ denote the density operator
for a closed
system at time $t$ (in Schr\H{o}dinger's picture)
and $U[t_i,t_f]$ denote the evolution
operator between $t_i$ and $t_f$, then $\rho(t_f)=
U[t_i,t_f]\rho(t_i)U[t_i,t_f]^{-1}$. The first of these
operators takes into account what will be called the
``forward time branch'' and the other one the
``backward time branch'' of the evolution. 
Evaluating matrix
elements of $\rho(t_f)$ can be done with an appropriate
propagator
which can be represented as a double functional integral
(one for each evolution operator). Correlation functions
are generated by introducing external sources
coupled
to the system's degrees of freedom which have
different values on the forward and backward time
branches \cite{Cugliandolo:1998-1}. Then, one usually takes a trace at time $t_f$. 
The generating functional obtained this way is called the
Keldysh generating functional and can be represented as a
path integral over a special contour $K$ that goes from $t_i$
to $t_f$ and then back to $t_i$.

Denoting by $\varphi$
(respectively $\xi$)
fields (respectively external sources) describing the dynamics of the system, and by
$\varphi_+$ and $\varphi_-$ their restrictions to the
upper and lower branch of Keldysh's contour
(respectively $\xi_+$, $\xi_-$), we have:
$$Z[\xi_+,\xi_-] = \int {\cal D}[\varphi_+,\varphi_-]\,
e^{i(S[\varphi_+,\xi_+]-S[\varphi_-,\xi_-])}\,\rho[\varphi_+|_{t_i},
\varphi_-|_{t_i}]
$$
where $\rho(t_i)[\varphi_+|_{t_i},
\varphi_-|_{t_i}]$ denotes the kernel of the initial
density operator. This is the basis for Keldysh's non
equilibrium perturbation theory.
Because of the doubling of degrees of
freedom, one obtains a 2 by 2 matrix Green function which can
be related to well known Green functions by:
\begin{equation}
\boldsymbol{G}(x,y) = \left(G_{\epsilon\epsilon
    '}(x,y)\right)_{(\epsilon,\epsilon ')\in \{1,-1\}} = 
\left(
\begin{array}{cc}
G_T(x,y) & G_{<}(x,y)\\
G_{>}(x,y) & G_{\tilde{T}}(x,y)
\end{array}\right)
\end{equation}
Here $T$ denotes the usual time ordering, $\tilde{T}$ the
anti-chronological time ordering, $<$ and $>$ the lesser
and upper time orderings \cite{Mahan}. For 
bosonic oscillators initially in thermal
equilibrium, Green's functions are explicitly known (see
appendix \ref{secAppGreen}).

\medskip

For a matter system coupled to QED, 
we are interested in the evolution of the reduced
density matrix for the matter system.
Bosonization techniques enable us to treat the electron
fluid through an effective bosonic theory. Our strategy will then
be to integrate over the electromagnetic field degrees of
freedom and to deal with matter degrees of freedom in a
second time. It is convenient to choose the
Coulomb gauge, in which dynamical degrees of
freedom of the quantum electromagnetic field (transverse
photons) are decoupled from the Coulomb interaction. The latter is
taken into account through the effective action for the
matter system which uses only
instantaneous interactions such as the Luttinger liquid description. 
Integrating out transverse photons gives a non-local
functional integral of the current density, called a
Feynman-Vernon influence functional \cite{Feynman:1963-1}.
For the sake of
simplicity, we shall assume that these two systems are
independent at initial time (the initial 
density operator
factorizes  $\rho(t_i) = \rho_{\mathrm{Matter}}\otimes
\rho_{\mathrm{QED}}$)
and that
the electromagnetic field is initially at equilibrium at
temperature $T=1/k_B\beta$.
In the present case, using bold symbols for vectors and matrices,
the influence functional is nothing but the 
influence functional for the current density:

\begin{equation}
\label{eqFV1}
{\cal F}[\boldsymbol{j}_+,\boldsymbol{j}_-] =
\int {\cal D}[\boldsymbol{A}_+,\boldsymbol{A}_-]\,
e^{i(S[\boldsymbol{A}_+]-S[\boldsymbol{A}_-])}
\times
e^{i\int
  (\boldsymbol{A}_+\boldsymbol{j}_+-\boldsymbol{A}_-\boldsymbol{j}_-)}\times
\rho_{\beta}[\boldsymbol{A}_+(t_i),\boldsymbol{A}_-(t_i)]
\end{equation}

where $\rho_{\beta}$ is the appropriate functional kernel
representing the density operator for the quantum
electromagnetic field at given temperature. 
The functional integral \ref{eqFV1} can be evaluated in
terms of
Keldysh's Green function for the electromagnetic field
at finite temperature
$D^{\alpha\beta}_{\epsilon\epsilon '}(x,y)=-i \langle T_K
A ^{\alpha}_{\epsilon}(x)A ^{\beta}_{\epsilon
  '}(y)\rangle$ 
and is equal to:
\begin{equation}
\label{eqFV2}
{\cal F}[\boldsymbol{j}_+,\boldsymbol{j}_-] =
\exp{\left(
-\frac{i}{2}\sum_{(\epsilon,\epsilon ')\in\{1,-1\}}
\epsilon\,\epsilon '\, \int d^4x\,d^4y \,
\boldsymbol{j}^{\alpha}_{\epsilon}(x)\,
D^{\alpha\beta}_{\epsilon\epsilon '}(x,y)\,
\boldsymbol{j}^{\beta}_{\epsilon '}(y)
\right)}
\end{equation}
Noise and dissipation kernels are defined by:
\begin{eqnarray}
\label{eqNoise}
\nu^{\alpha\beta}(x,y) & = & -\frac{1}{2}\,
\Im{(D^{\alpha\beta}_T(x,y))}\\
\label{eqDissip}
\eta^{\alpha\beta}(x,y) & = & -\frac{1}{2}\,
(D^{\alpha\beta}_R(x,y)-D^{\alpha\beta}_A(x,y))
\end{eqnarray}
and we have:
\begin{eqnarray}
\label{eqFV3}
{\cal F}[\boldsymbol{j}_+,\boldsymbol{j}_-] & = &
\exp{\left(
-\int d^4xd^4y\; {}^t(\boldsymbol{j}_+(x)-\boldsymbol{j}_-(x))
\ldotp\boldsymbol{\nu}(x,y)
\ldotp(\boldsymbol{j}_+(y)-\boldsymbol{j}_-(y))\right)}\\
& \times & \exp{\left(i\int d^4xd^4y\; {}^t(\boldsymbol{j}_+(x)-\boldsymbol{j}_-(x))
\ldotp\boldsymbol{\eta}(x,y)\ldotp
(\boldsymbol{j}_+(y)-\boldsymbol{j}_-(y))\right)}
\end{eqnarray}
Correlation functions can be obtained from a generating
functional which takes the environment into account:
$$\int {\cal D}[\varphi_+,\varphi_-]\;
e^{i(S[\varphi_+,\xi_+]-S[\varphi_-,\xi_-])}\,
{\cal
  F}[\boldsymbol{j}[\varphi_+],\boldsymbol{j}[\varphi_-]]
\,\rho(t_i)[\varphi_+(t_i),\varphi_-(t_i)]
$$
Within the elastic approximation, coupling to acoustic phonons can be treated along the
same line. Longitudinal
phonons create a potential $V(\boldsymbol{x},t)$ such that
$e\,V(\boldsymbol{x},t) =D \rho(\boldsymbol{x},t)
\mathrm{div}(\boldsymbol{u}(\boldsymbol{x},t))$ where 
$\boldsymbol{u}(\boldsymbol{x},t)$ is the elastic
deformation field, $\rho(\boldsymbol{x},t)$ is the
electric charge density and $D$ denotes the electron-phonon coupling
energy. Longitudinal phonon dynamics is described by a
quadratic action:
\begin{equation}
S[\boldsymbol{u}(\boldsymbol{x},t)]  = \frac{\rho_M}{2}\int 
\left(
(\partial_t\boldsymbol{u})^2 - c_S^2\,(\mathrm{div}(\boldsymbol{u}))^2
\right)
d^3\boldsymbol{x}dt
\end{equation}
where $c_S$ denotes the sound velocity and $\rho_M$ the
volumic mass. The
resulting phonon influence functional is a Gaussian in
terms of $\rho_{\pm}(x)$ with kernel 
$\mathcal{D}^{\alpha,\beta}_{\epsilon,\epsilon'}(x,y) = -i\langle 
\boldsymbol{u}^\alpha_{\epsilon}(x)\boldsymbol{u}^\beta_{\epsilon'}(y)
\rangle_{\beta}$:
\begin{equation}
\label{eqFV31}
{\cal F}[\rho_+,\rho_-] =
\exp{\left(
-\frac{iD^2}{2}\int d^4x\,d^4y\,
\sum_{(\epsilon,\epsilon')\in \{+,-\}^2}
\rho_{\epsilon}(x)
\rho_{\epsilon '}(y)
(\partial_{x,\alpha}\partial_{y,\beta}\mathcal{D}^{\alpha,\beta}_{\epsilon,\epsilon'})(x,y)
\right)}
\end{equation}

\medskip

In a generic electron system, this influence functional
{\it a priori} contains quartic fermion
terms. The one dimensional interacting electron gas \cite{Luttinger:1963-1} is quite
interesting since it can be described by an effective
{\em free bosonic theory}\cite{Haldane:1981-1}. Neglecting environment-induced {\em
  umklapp} processes, the influence functional is
Gaussian (although non local) in terms of the bosonic
field. Such terms arise from the introduction of $2k_F$
components of the charge or current densities in the
environment's influence functional. There is a factor
$\exp{(2ik_F(\sigma+\sigma'))}$ in front of the term involving 
two $\psi^{\dagger}_R\psi_L$s. At incommensurate filling,
the usual averaging argument can be invoked to rule out these {\em Umklapp}
contributions. Other terms contain both
$\psi^{\dagger}_R\psi_L$ and $\psi^{\dagger}_L\psi_R$ 
and would imply momentum transfer of order $2k_F$ (except for
special values of filling). In QED's case, $2k_F$ photons
would have a very high energy compared to the Luttinger
typical energy $\hbar v_S/L$. In the acoustic case,
since $k_FL>>v_S/c_S$, it would also be the case. That's
why we will not take into account $2k_F$ components of charge
and current densities. Thanks
to a Fourier mode analysis, this problem 
boils down to a set of independent harmonic
oscillators, each of them being linearly coupled to a
bath of quantum oscillators. This problem is known under
the name of {\em Quantum Brownian Motion} (QBM) and has widely
been studied \cite{CL:1983-1,CL:1983-2}. Necessary
results will be recalled in section \ref{secQBM}.
Elementary excitations of the Luttinger liquid can be
created using vertex operators. States created
this way are nothing but coherent states. Since the
evolution of coherent states in QBM can be exactly
computed \cite{CL:1985-1}, so can the evolution of elementary excitations
in the Luttinger liquid. 

\subsection{Mode decomposition for the Luttinger liquid}
\label{secLuttingerRecall}

Let us consider a Luttinger liquid on a
circle of length $L$, with both left and right
chiralities. As in 
\cite{Degiovanni:1998-1}, $\sigma$ denotes the coordinate
along the circle. Low energy excitations of the Luttinger liquid are described by 
the theory of a free compactified boson, 
whose action is given by
\begin{equation}
S[\varphi] = \frac{g}{2\pi}\int d\sigma\,dt\; 
(v_S^{-1}
(\partial_t\varphi)^2-v_S(\partial_{\sigma}\varphi)^2)
\end{equation}
The field $\varphi$ is compactified on a circle of radius $R_c$, and
the Euclidean path integral representing the finite temperature 
equilibrium partition function contains a topological
term taking into account zero mode quantization.
The space of states of
this free bosonic theory is a representation of a
$\widehat{U(1)}_R\times
\widehat{U(1)}_L$ algebra (two commuting 
copies of an oscillator algebra):
$[ J_k , J_l ] = k\,\delta _{k,-l}\,\boldsymbol{1}$ and 
$[ \bar{J}_k , \bar{J}_l ] = k\,\delta _{k,-l}\,\boldsymbol{1}$.
In Wen's work \cite{Wen:1990-1}, $J_l$ and $\bar{J}_l$ for
$l\neq 0$ are 
called {\em  hydrodynamic modes}, whereas the $J_0$ and $\bar{J}_0$
are called {\em zero modes}. 
In term of these modes, the Hamiltonian of the Luttinger system is 
given by:
\begin{equation}
\label{Hamil-J-inter}
H_{\mathrm{tot}} = \frac{\pi v_S}{L}
\sum _{l\in \mathbb{Z}}(
J_l J_{-l}+\overline{J}_l \overline{J}_{-l}).
\end{equation}
The finite temperature equilibrium partition function 
with a magnetic flux $\chi e/h$ is given by:
\begin{equation}
\label{eqRatio1}
Z_{\mathrm{Lutt}}(\chi)= {1\over |\eta (\tau)|^2}\,
\sum _{{(n,m)\in (\mathbb{Z}/2)\times \mathbb{Z}\atop
2n\equiv m\pmod{2}}}
q^{{1\over 2}p_{n,m+2\chi}^2}
\overline{q}^{{1\over 2}\overline{p}_{n,m+2\chi}^2},
\end{equation}
where $\alpha = gR_c^2$ encodes the interaction
strength ($\alpha = 1$ for free electrons) and
\begin{equation}
\label{eqRatio2}
p_{n,m}=n\sqrt{\alpha}+{m\over 2\sqrt{\alpha}}\quad
\overline{p}_{n,m}=n\sqrt{\alpha}-{m\over 2\sqrt{\alpha}}.
\end{equation}
States of interest will be $\widehat{U(1)}\times \widehat{U(1)}$ 
highest weight states:
with $J_0$ (resp. $\bar{J}_0$) 
eigenvalues $p_{n,m}$ (resp. $\bar{p}_{n,m}$). 
The charge and current densities can be expressed in
terms of the $\widehat{U(1)}$ modes as
follows:
\begin{eqnarray}
\label{eqDefQ}
\rho(\sigma,t) & = & {e\over L\sqrt{\alpha}}\, \sum _{n\in \mathbb{Z}}\left(
J_n(t)\, e^{2\pi i\, {n\over L}\,\sigma}
+\overline{J}_n(t)\, e^{-2\pi i\, {n\over L}\,\sigma}
\right)\\
\label{eqDefJ}
j(\sigma,t) & = & 
{ev_S\over L\sqrt{\alpha}}\, \sum _{n\in \mathbb{Z}}\left(
J_n(t)\, e^{2\pi i\, {n\over L}\,\sigma}
-\overline{J}_n(t)\, e^{-2\pi i\, {n\over L}\,\sigma}\right).
\end{eqnarray}
The $|n,m\rangle_{\chi}=|p_{n,m+2\chi},\bar{p}_{n,m+2\chi}\rangle$
state carries a charge $2n\in \mathbb{Z}$ 
and a current
$(m+2\chi)\,v_S/L\alpha$ \cite{Loss:1992-1,Degiovanni:1998-1}. 
If we assume that our Luttinger system 
is realized using the edge states of a FQH fluid, then the two edges
of the sample are polarized with charges ($\alpha =
1/\nu$ is an odd integer):
\begin{equation}
\label{eqSpecCharge}
q_R = e\,\left(n+{m\over 2\alpha}\right)\ \mathrm{and}\ 
q_L =  e\,\left(n-{m\over 2\alpha}\right).
\end{equation}
Obviously, $\widehat{U(1)}$ descendants carry the same global charges and
current than the highest weight states. Remember that
$\widehat{U(1)}$ primaries are created using vertex
operators $V_{n,m}(\sigma,t)$ which are
normal ordered exponentials of bosonic modes
\cite{Kac,FLMbis}. 
To complete this brief description, let us recall that the original fermionic
operators get renormalized (orthogonality catastrophe
historically introduced by Anderson \cite{Anderson:67-1}),
and that the corresponding renormalized fields are the
so-called {\em Luttinger ``fermions''} which correspond
to $n=\pm 1/2$ and $m=\pm 1$.
Within the context of fractional quantum Hall effect on a
cylinder -- or an annulus --
{\em edge fermions} carrying a
charge localized on one of the two edges, can be created
or destroyed using vertex operators \cite{Stone:94-1}
with $n=\pm 1/2$ and $m=\pm \nu^{-1}$.
Typical experiments which may be performed on edge excitations of an annular 2DEG in an
AsGaAl heterostructure lead to the following numerical
values which we shall use in the rest of this paper~: 
$v_S \simeq 3\ldotp 10^5
\mathrm{m}\,\mathrm{s}^{-1}$,  $L\simeq 30
\mu\mathrm{m}$, $\alpha =3$ and $T\simeq 0 -
10~K$. 

\subsection{Coupling of modes to the environments}

In order to respect cylindrical
geometry, the environment field is supposed to 
live in a cylindrical cavity with the same revolution
axis than the Luttinger circle.
The cavity is of radius $R$ and of height $h$ and
$\mathcal{V}=\pi R^2h$ will denote its volume. 
Periodic boundary conditions are assumed in this
vertical direction.

Electromagnetic modes are quantized
using 
their momentum along $z$ axis  $k_z$,
their angular momentum moment around $z$ axis $l\in \mathbb{Z}$
and the number $n$ of radial zeroes of the electric field.
The boundary condition on the cavity plays the role of
polarization and accounts for the helicity degeneracy.
Transverse electric ($TE$) modes have their (transverse) electric field
orthogonal to the cavity's edge, and transverse magnetic ($TM$)
modes have their magnetic
field orthogonal to the cavity's edge.
Normalized expressions for the modes can be found in any
book on optical fibers and are recalled in appendix
\ref{secAppModes}. Of course, the box is considered as
large meaning that electromagnetic modes form a
continuum compared to the Luttinger modes.

Phonons are quantized according to the vanishing of
displacement field $\boldsymbol{u}$ on the cavity's boundary.
Longitudinal modes can be written as a gradient
$\boldsymbol{u}=\boldsymbol{\nabla}(\varphi)$.
 As shown in
appendix \ref{secAppModes}, longitudinal modes are classified by 
$(k_z,l)$ and $n$ where $n$ is the number of zeroes of 
the $\varphi$ potential.

\medskip

It is convenient to introduce some kind of ``phase space
coordinates'' to
describe the $l\geq 1$ modes of the
Luttinger Liquid~:
\begin{align}
q_l^{(+)} &= \frac{1}{2\sqrt{l}} \big( J_l + J_{-l}
-\bar{J}_l - \bar{J}_{-l} \big)\quad \mathrm{and}\quad
i p_l^{(+)} = \frac{1}{2\sqrt{l}} \big( J_l - J_{-l}
-\bar{J}_l + \bar{J}_{-l} \big)\\
-i q_l^{(-)} &= \frac{1}{2\sqrt{l}} \big( J_l - J_{-l}
+\bar{J}_l - \bar{J}_{-l} \big) 
\quad \mathrm{and}\quad
p_l^{(-)} = \frac{1}{2\sqrt{l}} \big( J_l + J_{-l}
+\bar{J}_l + \bar{J}_{-l} \big)
\end{align}
Formulating the problem in this way
shows that the interaction Hamiltonian 
is nothing but a linear coupling
between harmonic oscillators ($qx$ coupling). 
More precisely, for each $l\geq 1$, two harmonic 
oscillators are coupled to the acoustic or
electromagnetic modes of angular momentum $l$. The two Luttinger modes
of fixed $l$ are coupled to different sets of environment modes. 
The same conclusion is true 
for zero modes: the global current couples to $l=0$ 
electromagnetic modes whereas the global charge decouples 
from the field's propagating modes (Coulomb gauge
effect). 
In the acoustic case, only the total electric charge
couples to $l=0$ acoustic modes.

\medskip

The  influence functional for each of
these environments can explicitly be obtained in a
nice form. First of all, the charge and current densities
can be expressed as linear combinations of the above phase
space coordinates.
Then, performing integration over the vector potential
and spatial coordinates leads to
the QED  functional (assuming 
that $q_0^{(+)}=q_0$ and $q_0^{(-)}=0$):
\begin{equation}
\label{eqQEDFV}
\mathcal{F}_{FV} = \exp{\left( -\frac{ig}{2}
\frac{L^3}{\mathcal{V}}\sum_{{I\atop l_I \geq 0}} \sum_{(\epsilon,
  \epsilon',\alpha)\in \{+,-\}^3} \epsilon\epsilon'\,
\frac{c^2}{L^2}\, \mathcal{D}_I^2l_I\,
\int\,dt\,ds\;q_{l_I}^{(\alpha)\epsilon}(t)\,
G_{\epsilon \epsilon'}^{(\omega_I)}(t-s)
\,q_{l_I}^{(\alpha)\epsilon'} (s)\right)}
\end{equation}
where the dimensionless coupling constant only depends on
the fine structure constant $\alpha_{QED}$, the Luttinger
liquid interaction parameter $\alpha$ and the ratio of
velocities $v_S/c$:
\begin{equation}
\label{eqCouplQED}
g=4\pi  \,\frac{\alpha _{QED}}{\alpha}\ldotp 
\left(\frac{v_S}{c}\right)^2
\end{equation}
This mode expansion is
related to a multipolar expansion.
Luttinger modes of momentum $l$ contribute to the
electric (resp. magnetic)
multipolar expansion starting at order $2l$ (resp. $2l+1$)
In particular, $l=0$ is mainly a magnetic dipole and
an electric monopole (corresponding to the global charge
and current around the Luttinger ring). The $l=1$ modes 
contribute first to the electric dipole and the magnetic
quadrupole.

\medskip

Longitudinal phonons can be treated in the same way. In
this case, one needs to perform a canonical
transformation on the Luttinger phase space coordinates,
exchanging $p_l^{(\pm)}$ and $q_l^{(\pm)}$ in order to
obtain a Gaussian expression similar to \ref{eqQEDFV}.
The dimensionless coupling constant
corresponding to longitudinal phonons in dimension $d=2,3$ is given by:
\begin{equation}
\label{eqCouplPhonons}
g_{ph}(L)={D^2\over \alpha\rho_ML^{d-1}\,\hbar\, c_S^3}
\end{equation}
where $c_S$ is the sound velocity, $\rho_M$ the volumic
($d=3$) or surfacic ($d=2$)
mass and $D$ the typical electron/phonon coupling
energy. With typical values
$D\simeq 7~eV$, $c_S\simeq 3\ldotp 10^3\
ms^{-1}$ and $\rho_M\simeq 3 \ldotp 10^3\ kg\, m^{-3}$,
we get, in dimension three $g_{ph}(L)\simeq 4\ldotp 10^{-6}$.

\section{Relation to the Quantum Brownian motion}
\label{secQBM}

\subsection{The Quantum Brownian Motion model}

This basic model consists in a single quantum harmonic
oscillator (called ``the system'') coupled to a bath of 
oscillators (called the ``reservoir''):
\begin{equation}
L = \frac{m}{2}\left(\dot{q}^2-\Omega^2q^2\right) + \sum_I C_I\,q\,q_I+
\sum_I\frac{M_I}{2}\left(\dot{q_I}^2-\omega_I^2q_I^2\right)
\end{equation}
We assume that these two systems are
initially independent and that the oscillator bath is at
equilibrium with inverse temperature $\beta$. 
One could also assume that the whole system is initially
at equilibrium but it makes things more involved without
illuminating the discussion. Interested readers are 
refered to Grabert's review \cite{Grabert:1988-1},
to Hakim and Ambegaokar's paper\cite{Hakim:1985-1} or
to Romero and Paz \cite{Romero:1996-1}. The
question addressed here is to compute the time evolution
of the system. The physics depends on the reservoir's
influence at each frequency, encoded in 
appropriate {\em spectral densities}. Therefore, several natural
time scales will appear in the Luttinger \& environment problem:
%
%
\newcommand{\taulutt}{\tau_L}
\newcommand{\taulight}{\tau_{EM}}
\newcommand{\tauthermal}{\tau_{Th}}

\begin{itemize}

\item The {\bf cut-off} time~: above a certain UV
  cutoff, the effective
  description by a Luttinger liquid is no longer
  valid. Another natural cutoff could also be provided by the
  spectral distribution itself (Debye frequency for
  phonons). For this reason, an UV cutoff is present in
  the model.
The associated time will be denoted by $1/\Lambda$. Below this
  time scale, our simple model cannot be considered as valid.

\item The {\bf environment time scale $\tau_E$} is
  the time needed by the light ($\taulight= L/c$) or by
  phonons ($\tau_S=L/c_S$) to circle around the
  system. 

Spectral densities and 
influence functionals
can be normalized with respect to this characteristic
time (they only depend on $\omega\,\tau_E$). 
The low frequency regime is defined
by $\omega\tau_E<<1$ and the high frequency regime by
$\omega\tau_E>>1$.

\item The {\bf Luttinger time scale} $\tau _L=L/v_S$ is
  the time needed by one excitation of the Luttinger
  liquid to circle around the system. This is the natural time scale from the Luttinger 
liquid point of view.

\item The {\bf thermal time scale} $\tauthermal=\hbar
  /k_BT$
is the inverse frequency associated with
temperature $T$. The corresponding thermal length is
$l(\beta)=c\,\tauthermal$ for photons and
$l(\beta)=c_S\tauthermal$ for phonons. Let us notice
that at $T\simeq \ 1\,\mathrm{K}$, 
$l(\beta)=1~\mathrm{mm}$ for photons whereas it is of order 10~$\mu \mathrm{m}$
for phonons.

\end{itemize}

The general
solution to the Quantum Brownian Motion problem is due to
Hu, Paz and Zhang (HPZ) \cite{HPZ:1992-1}
who computed the evolution kernel for the system's
reduced density matrix. Caldeira and Leggett's work
\cite{CL:1983-2} is
concerned with a special case of reservoir, corresponding
to the so-called 
Ohmic spectral density $\mathcal{J}(\omega)\propto \omega$.
As we have already noticed, the Luttinger liquid can be
seen as a collection of harmonic oscillators and some
zero modes. As we shall see, each harmonic mode coupled
to the environment is a QBM problem. 
Each reservoir is characterized by
a set of spectral functions (one for each $l\geq 0$), which are made dimensionless
for simplicity.

\medskip

In the electromagnetic case, using expressions  \ref{eqCoeffDTE}
and \ref{eqCoeffDTM} in appendix \ref{secAppModes}, we obtain:
\begin{equation}
\label{eqSpecDens}
\mathcal{J}_l (\omega) = g\,\frac{L^3}{\mathcal{V}}
\sum_{I\,/\ l_I=l} 
\frac{l_I\mathcal{D}_I^2}{2\,\omega_I\taulight}\, \delta(\taulight(\omega - \omega_I))
\end{equation}
The $l=0$ case needs a slight modification:
\begin{equation}
\label{SpecDens}
\mathcal{J}_0 (\omega) = g\,\frac{L^3}{\mathcal{V}}
\sum_{{I\in TE\atop l_I=0}} 
\frac{\mathcal{D}_I^2}{2\,\omega_I\taulight}\, \delta(\taulight(\omega - \omega_I))
\end{equation}
The electromagnetic influence functional can be
rewritten as:
\begin{multline}
\label{eqFeynmanVernonLutt}
\mathcal{F}_{FV}[\boldsymbol{j}_+,\boldsymbol{j}_-] 
= \exp{\left( i{c^2\over L^2}\sum_{l=0}^{\infty}
\sum_{\alpha=+,-} \int_{t_i}^{t_f} dt \int_{t_i}^{t} ds\,
(q_l^{(\alpha)+}-q_l^{(\alpha)-})(t)\,\eta_l(t-s)\,
(q_l^{(\alpha)+}+q_l^{(\alpha)-})(s)\right)} \\
\times \exp{\left( -{c^2\over L^2}\sum_{l=0}^{\infty}
\sum_{\alpha=+,-} \int_{t_i}^{t_f} dt \int_{t_i}^{t} ds\,
(q_l^{(\alpha)+}-q_l^{(\alpha)-})(t)\,\nu_l(t-s)\,
(q_l^{(\alpha)+}-q_l^{(\alpha)-})(s)\right)}
\end{multline}
Dimensionless 
dissipation and noise kernels for each
mode are expressed in terms of the spectral density by:
\begin{align}
\eta_l (s) &= -\frac{L}{c} \int_0^\infty d\omega\,
\mathcal{J}_l (\omega) \sin (\omega s)
\\
\nu_l (s) &= \frac{L}{c} \int_0^\infty d\omega \,\mathcal{J}_l
(\omega)  \coth {\left(\frac{\beta \omega}{2}\right)}
\cos(\omega s)
\end{align}
An acoustic reservoir leads to similar expressions for 
spectral functions (using expressions \ref{eqNCoeff2} and
\ref{eqNCoeff3}):
\begin{equation}
\label{eqSpecDensAcoustic}
\mathcal{J}_l (\omega) = g_{Ph}(L)\,\frac{L^3}{\mathcal{V}}
\sum_{I\,/\ l_I=l} 
l_I\mathcal{N}_I^2\,(\tau_S\omega_I)^3\, \delta(\tau_S(\omega - \omega_I))
\end{equation}
The $l=0$ case needs a slight modification:
\begin{equation}
\label{SpecDensAcoustic}
\mathcal{J}_0 (\omega) = g_{Ph}(L)\,\frac{L^3}{\mathcal{V}}
\sum_{I\,/\ l_I=0} 
\mathcal{N}_I^2\,(\tau_S\omega_I)^3\, \delta(\tau_S(\omega - \omega_I))
\end{equation}

\medskip

Numerical computations of spectral densities as well
as analytic estimates of their asymptotics are available
(see appendix \ref{secAsymptotic}). The electromagnetic
case is illustrated on figure 3. 
In this case, all $l\neq 1$ modes are supraohmic
($\mathcal{J}_l(\omega)$ decreases faster than $\omega$) at
low frequency.
The $l=1$ modes show an ohmic behavior
($\mathcal{J}_1(\omega)\propto \omega$).
In this case, the dissipation kernel is local in time
and, as
we shall see in section \ref{secCaldeira}, an effective
Caldeira-Leggett model can be used to perform analytic
computations. 


In the acoustic case $\mathcal{J}_l(\omega)$ goes as
$(\omega\tau_S)^{2l+d}$ in the low frequency regime and
as $(\omega\tau_S)^{d-1}$ at higher frequencies. The main
difference between the electromagnetic and acoustic
reservoirs is that the natural Luttinger frequency $2\pi
/\taulutt$ falls in the low frequency domain for QED,
whereas it does not for phonons since
$\tau_S/\taulutt>>1$.

\subsection{Phase space evolution using Wigner functions}
\label{secDecoh}

The time evolution can be computed using the Wigner
function associated with the system's density
operator. This form is especially adapted to the study of
decoherence of Gaussian wave packets. Moreover, it
provides a nice quasi-classical insight on the evolution
of the system since in the Luttinger liquid, charge and current density
fluctuations play the role of ``phase space coordinates''
for the hydrodynamic modes and within the FQH effect framework,
encode the shape of the incompressible quantum Hall fluid
droplet \cite{Wen:1990-1}. The Wigner
function associated with an operator $B$ is defined by:
\begin{equation}
W_B(p,q)=\int dy\; e^{ipy/\hbar}\,\langle
q-\frac{y}{2}|B|q+\frac{y}{2}\rangle
\end{equation}
We use the following notation for phase space: 
$\phi=\left(
\begin{array}{c}
p \\ 
q
\end{array}
\right)$. The evolution kernel for the Wigner function
can be computed from the density operator evolution
kernel and is given by:
\begin{equation}
J_W(p,q,t|p_0,q_0,0)= {\cal N}(t)\,\exp{
\left(-\frac{1}{4}{}^t(N\ldotp
    \phi + N_0\ldotp \phi_0)\ldotp A(t)\ldotp (N\ldotp \phi
    +N_0\phi_0)\right)}
\end{equation}
where  the $N_0$ and $N$ time dependent matrices are given by:
\begin{equation}
\label{eqDefNmatrices}
N=\left(
\begin{array}{cc}
-1 & \dot{u}_2(t)\\
0 & -\dot{u}_2(0)
\end{array}
\right)\quad\mathrm{and}\ 
N_0=\left(
\begin{array}{cc}
0 & \dot{u}_1(t)\\
1 & -\dot{u}_1(0)
\end{array}
\right)
\end{equation}
The
$u_i(t)$ ($i\in \{1,2\}$) functions are defined in Hu,
Paz and Zhang's paper \cite{HPZ:1992-1} as solutions to the classical
equations of motion with dissipation and boundary
conditions $(u_1(0),u_1(t))=(1,0)$ and
$(u_2(0),u_2(t))=(0,1)$.
The $A(t)=(a_{i,j}(t))_{(i,j)\in \{1,2\}^2}$ matrix is
defined by 
$$a_{i,j}(t) = \frac{1}{2}\int_0^tds_1\int_0^tds_2\,
u_i(s_1)\,\nu(s_1-s_2)\,u_j(s_2).$$
The reader can check that, when the reservoir decouples, 
the evolution kernel
reduces to a delta function, giving back the
classical evolution in phase space.
Henceforth, we shall call $-N^{-1}N_0$ the ``Wigner
evolution operator'' and denote it by $U_t$. When turning on the coupling to the
environment, the classical delta distribution in phase
space spreads into a Gaussian, the center of which moves
according to $U_t$. As we shall now see, the evolution kernel
encodes all effects of dissipation and decoherence.

\medskip

The evolution of a density operator built from Gaussian
wave packets can be computed exactly. The generic form of a Gaussian
Wigner function is,
up to some normalization:
\begin{equation}
\label{eqWignerGeneric}
{\cal W}[\bar{\phi},K,Q]\,(\phi) = \exp{\left(-\frac{1}{2}
{}^t(\phi - \bar{\phi})\ldotp Q\ldotp (\phi-\bar{\phi})+i{}^tK\ldotp\phi
\right)}
\end{equation}
The 2 by 2 matrix $Q$ encodes the spreading of the
packet, $\bar{\phi}$ is the center of the packet and $K$
a phase modulation. Gaussian wave packets and in particular coherent states 
lead to Gaussian Wigner functions. A
Gaussian Wigner function remains Gaussian. 
If $W(t=0)={\cal W}[\phi_0,K_0,Q_0]$
then, after time $t$, 
\begin{equation}
W(t)= e^{-d[K_0,Q_0](t)}\times {\cal W}[U_t\ldotp \phi_0,K(t),Q(t)]
\end{equation}
where the basic parameters at time $t$ are given by
$d[K_0,Q_0](t) = {}^tK_0\ldotp D(t)\ldotp K_0$ and:
\begin{eqnarray}
 D(t) & = & \left(2Q_0+{}^tN_0A(t)^{-1}N_0\right)^{-1}
\label{eqDecohCoeff}\\
K(t) & = & {}^tU_t^{-1}\ldotp
(\boldsymbol{1}-2\,Q_0\,D(t))\ldotp K_0\\
Q(t) & = & {}^tU_t^{-1}\ldotp \left(
Q_0-2\,Q_0\,D(t)\,Q_0
\right)\ldotp U_t^{-1}
\end{eqnarray}
The $D(t)$ matrix contains the decoherence
effect. To understand this, let
us start with a coherent superposition of two
Gaussian wave packets $(\psi_1+\psi_2)/\sqrt{2}$. The
initial Wigner function is given by:
$$W(0)=\frac{1}{2}\left(W_{11}+W_{12}+W_{21}+W_{22}\right)
\quad \mathrm{where}\ 
W_{\alpha\beta}(0)=W[\bar{\phi}_\alpha,\bar{\phi}_{\beta},Q_0]={\cal
  W}\left[
\frac{\bar{\phi}_{\alpha}+\bar{\phi}_{\beta}}{2},
-i\sigma^y\ldotp(\bar{\phi}_\alpha-\bar{\phi}_{\beta}),Q_0
\right]$$
Under time evolution, the form of the wave packet is
preserved up to a global factor and a phase modulation:
\begin{equation}
W_{\alpha\beta}(t)=e^{-d_{\alpha\beta}(t)+i\Theta_{\alpha\beta}(t,\phi)}\;
W[U_t\ldotp\bar{\phi}_\alpha,U_t\ldotp\bar{\phi}_{\beta},Q(t)]
\end{equation}
where:
\begin{eqnarray}
d_{\alpha\beta}(t) & = &
d[-i\sigma^y\ldotp(\phi_\alpha-\phi_\beta),Q_0] 
\label{eqDecoh2} \\
\label{eqPhase}
\Theta_{\alpha\beta}(t,\phi) & = & {}^tK_q(t)\ldotp
(\phi-\bar{\phi}_{12}(t))\\
K_q(t) & = & K(t) -
\frac{\dot{u}_1(t)}{\dot{u}_2(0)}\,{}^tU_t^{-1}\ldotp
K_0
\end{eqnarray}
Each density operator $|\psi_\alpha\rangle\langle\psi_\alpha|$ has
its own evolution, described by 
$\bar{\phi}_\alpha(t)=U_t\ldotp\bar{\phi}_{\alpha}$ and
$Q(t)$. Note that $d_{11}(t)=d_{22}(t)=0$. 
The coherence part is contained in $W_{12}$, and
evolves according to $U_t\ldotp \bar{\phi}_{12}$ and
$Q(t)$ plus an exponential factor 
$e^{-d_{12}(t)+i\Theta_{12}(t,\phi)}$. The $d_{12}(t)$ factor gives the
attenuation of off-diagonal correlations and should
therefore be interpreted as the {\em decoherence
factor} between the two wave packets.

\medskip

Unfortunately, as noticed before, explicit and closed expressions for $U_t$ 
and $D(t)$ are not known for a general supra-ohmic environment. Therefore,
we shall perform a perturbative expansion in the
coupling constant. In the case of the decoherence factor
$d_{12}(t)$, it is enough to start from standard
harmonic oscillators expressions for the $u_1$ and $u_2$ functions
since $A(t)$ is of first order in the coupling constant. The
decoherence matrix can then be expressed as an integral
over the spectral density:
\begin{equation}
\label{eqDecohPerturbative}
D(t)  = \int _0^{+\infty} {\cal J}(\omega)\,
D(\Omega,\omega,t)\,\coth{\left(\frac{\beta\hbar \omega}{2}\right)}\,
d\omega.
\end{equation}
where $D(\Omega,\omega,t)$ has an explicit but involved expression.
Let us discuss its asymptotics in various physically relevant limits:

\begin{itemize}

\item At short times $\Omega t<<1$,
$\Omega$ can be neglected provided the spectral
density of the bath contains modes with frequencies much
higher than $\Omega$. Then, this regime is dominated
by the high frequencies of the bath and:
\begin{equation}
D(\Omega,\omega,t)
\simeq \left(
\begin{array}{cc}
\frac{t^2}{2} &
\frac{\omega t^3}{4}\\
\frac{\omega t^3}{4}&
\frac{t^4}{8}
\end{array}
\right)
\end{equation}
The last expression is only valid at very short times and
contains non-Markovian effects. Even in the
Caldeira-Leggett model, at very short times, all
frequencies of the bath take part in the evolution of the
system, leading to these non-Markovian effects.

\item When the condition $\Omega t<<1$ is no
  longer valid, dissipation effects
  with exponential relaxation will appear through linear
  terms in $t$. We must of course assume that $t$ is much
  smaller than typical relaxation times.
At growing times, linear terms will dominate
oscillating resonant ones and provide
a ``Golden rule'' estimate for the relaxation and decoherence times.
This approximation will be used in the next section in order
to evaluate the typical decoherence time of
Schr\H{o}dinger cat states in the Luttinger model.

\end{itemize}

When $t$ reaches the dissipation time scale, this
perturbative treatment breaks down. The long time regime
of decoherence can however be computed in some cases using
the Caldeira-Leggett model.

\section{Mutual decoherence of elementary excitations in a
 Luttinger liquid}
\label{secDecLL}

\subsection{Statement of the problem}

As recalled in section \ref{secLuttingerRecall}, elementary excitations of the
Luttinger liquid are created by vertex operators 
$V_{n,m}(\sigma)$.
Such a state is {\em not} an eigenstate of the
Hamiltonian but rather corresponds to the introduction of
a ``localized'' excitation at point $\sigma$ around the circle. 
Let us consider a Schr\H{o}dinger cat state build as a superposition of a
right moving Luttinger fermion ($n=1/2$
and $m=1$)
at different places around the circle:
\begin{equation}
\label{defCatRR}
|\psi_{RR}(0)\rangle = \frac{1}{\sqrt{2}}\left(
\psi_R^{\dagger}(\sigma_1)|0\rangle 
+ \psi_R^{\dagger}(\sigma_2)|0\rangle 
\right)
\end{equation}
In an isolated system, such a state will evolve according to:
\begin{equation}
|\psi_{RR}(t)\rangle = \frac{1}{\sqrt{2}}\left(
\psi_R^{\dagger}(\sigma_1,t)|0\rangle 
+ \psi_R^{\dagger}(\sigma_2,t)|0\rangle 
\right)
\end{equation}
Therefore, this coherent superposition will remain
coherent, as it should in any isolated quantum system. 
Switching on the coupling to the quantum electromagnetic
field or phonons changes the situation: according to general works
on decoherence
\cite{Unruh:1989-1,Zurek:1982-1,Unruh:1989-1,Zurek:1991-1},
we expect this
Schr\H{o}dinger cat to decohere into a
statistical mixing of two states: one excitation at one
position, or the excitation at the other position. 
Here, two questions will be addressed: what is the strength of the decoherence
process and on which time scale does it take place?

\medskip

In the following, two cases will be considered: 
the $R/R$ Schr\H{o}dinger cat, already presented in
equation \ref{defCatRR} and the $R/L$ Schr\H{o}dinger cat, defined as:
\begin{equation}
\label{defCatRL}
|\psi_{RL}(0)\rangle = \frac{1}{\sqrt{2}}\left(
\psi_R^{\dagger}(\sigma_1)|0\rangle 
+ \psi_L^{\dagger}(\sigma_2)|0\rangle 
\right)
\end{equation}
Practically, the case of zero modes is simpler and will
be considered in the next paragraph. 
We shall then turn to the $l\neq 0$ modes in
section \ref{secNonzeroModes}. Decoherence time estimates
for electromagnetic ({\em resp.} acoustic) reservoirs are
given in \ref{secQEDTime} ({\em resp.} \ref{secAcousticTime}).
An effective
Caldeira-Leggett model will be used to deal with the
long time behavior in sections \ref{secCaldeira} to
\ref{secLongTime}.

\subsection{Evolution of zero modes}
\label{secZeroModes}

For the zero modes, the zero coupling evolution
corresponds to a free particle (and {\em not} to an
harmonic oscillator).
The
Ambegaokar and Hakim \cite{Hakim:1985-1} 
method (exact diagonalization for the coupled
system) can be used to compute the evolution of the
density matrix for the zero modes.
The corresponding explicit formula in the Luttinger
liquid case is, at finite temperature:
\begin{equation}
\label{eqZeroM}
\langle n,m|\rho(t)|n',m'\rangle = 
\langle n,m|\rho(0)|n',m'\rangle 
\times \exp{\left( 
-i(\omega_{n,m}(t)-\omega_{n,m}(t))\, t
\right)}\times 
\exp{\left( -\frac{d(t)}{\alpha}(m-m')^2 \right)}
\end{equation}
where $\hbar\omega_{n,m}(0)$ denotes the energy of $|n,m\rangle $ 
in the isolated Luttinger system and:
\begin{eqnarray}
\label{eqFreqRenorm}
\omega_{n,m}(t) & = &
\omega_{m,n}(0) + \frac{\pi v_S}{2L\alpha}\,m^2\,
\int_0^{+\infty} \frac{c}{\pi v_S}
\mathcal{J}_0(\omega)\,
\left(\frac{\sin(\omega t)}{\omega t}
  -1\right)\frac{d\omega}{\omega}\\
d(t) & = & \frac{t^2}{\taulight}
\int_0^{+\infty} d\omega \,\mathcal{J}_0(\omega)\,
\coth{\left({\beta\hbar\omega\over 2}\right)}\,
\frac{1-\cos(\omega t)}{(\omega t)^2}
\label{eqDecohExp}
\end{eqnarray}
The first term in \ref{eqZeroM} 
can be interpreted as a dynamical renormalization
of $v_S/\alpha$. Only $v_S/\alpha$ is
renormalized since the charge density does not couple to
the transverse degrees of freedom of the electromagnetic field.
To be precise, at $t\rightarrow +\infty$, the velocity and interacting 
parameters of the Luttinger liquid get renormalized as
$v'_S = v_S\ldotp \zeta$ and
$\alpha ' = \alpha / \zeta$ where the renormalization constant 
$\zeta$ is equal to the $t\rightarrow +\infty$ limit of:
\begin{equation}
\label{eqRenorm}
\zeta(t) ^2 = 1+\frac{c}{\pi v_S}\,\int_0^{+\infty}
\mathcal{J}_0(\omega)\,\left(\frac{\sin(\omega t)}{\omega t}
  -1\right)\,\frac{d\omega}{\omega}
\end{equation}
The renormalization effect is of course strongly cut-off dependent. The 
dimensionless coupling constant appearing here is 
$\frac{\alpha_{QED}}{\alpha}\ldotp \frac{v_S}{c}\simeq 10^{-5}$.

\medskip

The second term is the decoherence coefficient between two different
highest weight states $|n,m\rangle$ and $|n',m'\rangle$
of the LL. Decoherence takes place in a time of
the order of the cutoff time $\Lambda^{-1}$ and then
reaches saturation. Figure 4 summarizes
$d(t)$ and $\zeta(t)$'s behavior.


The typical value of the $t\rightarrow
+\infty$ value of the decoherence exponent is of typical order $g$:
\begin{equation}
\label{eqd0infini}
d(+\infty)=\int _0^{+\infty}
{c\over L\omega^2}\,\mathcal{J}_0(\omega)\,
\coth{\left(\frac{\beta\hbar\omega}{2}\right)}\,d\omega
\end{equation}

In the acoustic case, computations can be performed in
the same way. Since acoustic zero modes couple to the
total charge of the Luttinger system, the decoherence
factor between states $|n,m\rangle$ and $|n',m'\rangle$
is found to be $\exp{(-2\,\alpha\,d(t)(n^2-(n')^2))}$
where $d(t)$ is obtained from \ref{eqDecohExp}
by using the acoustic spectral density and
$\tau_S$ insted of their electromagnetic counterparts. The Luttinger parameters
$\alpha $ and $v_S$ also get renormalized. In the
acoustic case, only $v_S\alpha $ is renormalized. Then, 
$\alpha'=\alpha\zeta_{ph}$ and $v_S'=v_S\zeta_{ph}$ where
$\zeta_{ph}$ is the $t\rightarrow +\infty$ limit of
$\zeta_{ph}(t)$, obtained by using the acoustic spectral
density and the speed of sound in formula \ref{eqRenorm}.

\medskip

In both cases, the final decoherence exponent is proportional to
the square of the difference between the total current
({\it resp.} charge),
quantities which measure the ``distance'' between the
two quantum states. Such a
result is expected since, as explained in
C.~Cohen~Tannoudji's lectures \cite{Cohen:CDF:1989}, 
such a dependence is common in the case of a linear coupling 
with a conserved quantity.
In particular, a Schr\H{o}dinger cat obtained by superposing
the {\em same} elementary excitation of the Luttinger
liquid at two different positions along the ring has all
its decoherence due to hydrodynamic modes~! 
We also notive that zero mode decoherence has a weak
dependence in the cut off and temperature. 

\subsection{Spatial dependence of decoherence}
\label{secNonzeroModes}

Using the explicit form of vertex operators, one easily
finds the relevant parameters to be used for the
decoherence of each mode. Of
course, these parameters depend on positions of each
of vertex operator. Here, we shall only present the
results for $R/R$ and $R/L$ Schr\H{o}dinger cats
($\sigma_{12}=\sigma_1-\sigma_2$):
\begin{align}
d_{RR}(t,\sigma_1,\sigma_2) &= 
\frac{4}{\alpha}
\sum _{l=1}^{+\infty}
\frac{1}{l}\,
(m^2D_{11}^{(l)}(t)+4n^2\alpha^2D_{22}^{(l)}(t))
\,\sin^2{\left(\frac{\pi l\sigma_{12}}{L}\right)}\\
d_{RL}(t,\sigma_1,\sigma_2) &= 
\frac{4}{\alpha}
\sum _{l=1}^{+\infty}
\frac{1}{l}\,
\left\{
m^2D_{11}^{(l)}(t)\left(
1-\sin^2{\left(
\frac{\pi l\sigma_{12}}{L}
\right)}
\right)\right.\\
 &+ \left. 4n^2\alpha^2D_{22}^{(l)}(t)\,
\sin^2{\left(\frac{\pi l\sigma_{12}}{L}\right)}
+2nm\,\alpha D_{12}^{(l)}(t)\,
\sin{\left(\frac{2\pi l\sigma_{12}}{L}\right)}
\right\}
\end{align}
Here $D^{(l)}(t)$ denotes the decoherence matrix for the $q_l^{(\pm)}$
modes computed along the lines of section
\ref{secDecoh}. The main change from the HPZ computations
arises from our normalization choice for the $q_l$s. The effective
spectral density to be used in the HPZ formulas is given by: 
$$2\pi l\,{v_S\over c}\ldotp{\mathcal{J}_l(\omega)\over
\tau_E^2}.$$
This rescaling takes into account the ratio
of the Luttinger mode $l$ time scale
({\it i.e.} $\tau_L/l$) and of the environment time
scale $\tau_E=L/c$.

\medskip

The appearance of an odd dependence -- in term of
$\sigma_{12}$ --  in the
$d_{RL}(t,\sigma_1,\sigma_2)$ coefficient is understood
by noticing that an appropriate parity operation
transforms the $R/L$ Schr\H{o}dinger cat into an $L/R$
one. Therefore $d_{RL}(t,\sigma_1,\sigma_2)$ is invariant
into simultaneous changes $\sigma_1\leftrightarrow \sigma_2$ and $nm\mapsto -nm$.

\medskip

A first estimate is obtained using a perturbative
approach. We perform a secular
approximation and retain only terms linear in time. The
corresponding decoherence rates are given by:
\begin{eqnarray}
\label{eqDecohRate}
{\tau_L\over
  \tau_l^{(R/R)}(\sigma_1,\sigma_2)} & = &
8\pi^2\Delta_{n,m}\,\,
\ldotp\,\sin^2{\left( {l\pi\sigma_{12}\over L}
  \right)}\,\,
{\mathcal{J}_l(\omega_l)\over \omega_l\tau_E}\,\\
{\tau_L\over
  \tau_l^{(R/L)}(\sigma_1,\sigma_2)} =  & = & 
8\pi^2\,\Delta_{n,m}
\,\ldotp\left(
1+{m^2-4n^2\alpha^2\over
  m^2+4n^2\alpha^2}\,
\cos{\left({2\pi l\sigma_{12}\over L}\right)}\right)
\,{\mathcal{J}_l(\omega_l)\over \omega_l\tau_E}
\end{eqnarray}

Here $\Delta_{n,m}$ is the conformal dimension of the
vertex operator $V_{n,m}(\sigma)$. Not surprisingly, the decoherence time of a $R/R$
Schr\H{o}dinger cat diverges when
$\sigma_{12}\rightarrow 0$. This result is
obvious since in this limit, the initial state is a pure
state. For $L/R$ cats, the decoherence time shows a slow
variation in term of the differences of positions. 

\subsection{Decoherence time estimations: QED's case}
\label{secQEDTime}

Using asymptotics of spectral densities (see appendix \ref{secAsymptotic}), 
one obtains the decoherence time of the $l$th modes
in the $R/R$ case:
\begin{equation}
{\tau_L\over
  \tau_l^{(R/R)}(\sigma_1,\sigma_2)} =  4 g\,
\Delta_{n,m}\left({2\pi v_S\over c}\right)^{2(l-1)}\ldotp 
{l^2(l+1)\over (2l+1)!}\,\ldotp
\sin^2{\left({\pi l\sigma_{12}\over L}\right)}
\end{equation}
Similarly, for the $R/L$ case:
\begin{equation}
{\tau_L\over
  \tau_l^{(R/L)}(\sigma_1,\sigma_2)} 
= 4g\, \Delta_{n,m}\left({2\pi v_S\over c}\right)^{2(l-1)}\ldotp 
{l^2(l+1)\over (2l+1)!}\,\ldotp 
\left(1+{m^2-4\alpha^2 n^2
\over m^2+4\alpha^2 n^2}
\cos{\left({2\pi l\sigma_{12}\over L}\right)}
\right)
\end{equation} 
Since $v_S/c\simeq 10^{-3}$, decoherence times for the $l$ and $l+1$ 
modes are related by a typical factor of $10^6$. This
argument shows that the $l=1$ modes dominate
the decoherence process. Physically, higher Luttinger
modes contribute to higher electric and magnetic multipoles,
for which radiative dissipation is known to be
weaker. Since dissipation governs decoherence, 
this is the physical reason for the predominance
of the $l=1$ Luttinger mode in the decoherence process.
The decoherence time
of the $l=1$ mode is nothing but
the electromagnetic relaxation time:
$$\tau^{-1} \simeq {16\pi
\over 3}\,{\alpha_{\mathrm{QED}}\over \alpha}
\,\left({v_S\over c}\right)^2\ldotp\tau_L\simeq 10^{-8}\,\tau_L^{-1}$$
Numerical results for the decoherence times are shown on figure
5 for Luttinger fermions. 


The temperature dependence can be found easily
since $\coth{(\beta\hbar \omega/2)}$ varies slowly
around $\omega_l$ in a scale $g\omega_l$. 
Therefore:
\begin{equation}
\frac{\tau_l^{(R/R)}(\sigma_1,\sigma_2,T)}{\tau_l^{(R/R)}(\sigma_1,\sigma_2,T=0)}
=\tanh{\left(\frac{\hbar\,\omega_l}{2k_B\,T}\right)}
\end{equation}

\subsection{Decoherence time estimations: acoustic case}
\label{secAcousticTime}

Since the sound velocity $c_S$ is much smaller than 
$v_S$, the condition $\omega_l\tau_E<<1$ 
does not hold for the coupling to phonons. 
Explicit computations show that the behavior
of spectral densities for the coupling to longitudinal
phonons differs from the electronic one.
Luttinger modes frequencies fall into a range of frequencies where the 
acoustic spectral densities are proportional to
$(\omega \tau_S)^{d-1}$. Remember also that the natural
cut-off frequency for the phonon bath is given by
$\omega_D=c_S/a$ where $a$ is a typical
microscopic length\footnote{One would naturally think of
  $\omega_D$ as the Debye frequency but a
  linear dispersion relation for phonons has been
  assumed, an assumption which certainly fails near the
  Debye frequency. The $a$ length scale should rather be
  considered as small compared to the mesoscopic size of
  the system but large compared to the atomic length scale.}.
The zero temperature
acoustic decoherence rate of Luttinger modes is given by
formulas \ref{eqDecohRate} where the $\Gamma _l^{(d)}=\mathcal{J}_l(\omega_l)/
  \omega_l\tau_S$ factor is given by:
\begin{eqnarray}
\Gamma _l^{(3)} & = &  \frac{g_{ph}(L)\,l^2}{2}\,\frac{v_S}{c_S}\\
\Gamma _l^{(2)} & = &  \frac{g_{ph}(L)\,l}{\pi}
\end{eqnarray}
In opposition to QED's case, these damping rates do not
decrease with increasing $l$. In the QED
case, higher $l$ modes are bad antennas for
the microwave radiation emitted by the system. In the
case of phonons the situation goes the other way because
of the longitudinal coupling.
Although in some cases the coupling 
constant $g_{ph}(L)$ is very small, 
``decoherence repartition'' effects between modes plays a much more
important role here than in QED's case since
one has to sum up over many mode contributions to decoherence. 
For Luttinger modes of index $l$, the 
perturbative expansion is governed by the relative
damping rate $\gamma _l/\omega_l$, an upper value of which is given by 
\begin{equation}
\frac{g_{ph}(a)}{8\pi^2}\, \frac{\omega_l\tau_S}{(\omega_D\tau_S)^2}
\quad \mathrm{for}\ d=3\quad \mathrm{and}
\quad \frac{g_{ph}(a)}{\pi\,(\omega_D\tau_S)}\ \mathrm{for}\ d=2
\end{equation}
where $g_{ph}(a)$ is the rescaled coupling constant for the
length $a$ (typically of order $1$). Assuming that $\omega_D\tau_S=L/a$ is much
greater than one, we see that {\em all $l\neq 0$ modes
can be considered as weakly damped}.

\medskip

The total decoherence exponent in the linear regime is
obtained by summing over all the modes up to the Debye frequency. 
For $R/R$ Schr\H{o}dinger cat states, one finds:
\begin{equation}
\tau_L\,\ldotp \gamma^{(R/R)} (\sigma_1,\sigma_2,T) = 
8\pi^2\Delta_{n,m}\,\sum_{l=1}^{l_{max}}
\Gamma ^{(d)}_l\,\coth{\left( \frac{\beta\hbar\omega_l}{2}\right)}\,
\sin^2{\left(\frac{\pi \, l \, \sigma_{12}}{L}\right)}
\end{equation}
Since we sum over a large number of modes, the
decoherence time rapidly decreases when
$\sigma_{12}>>av_S/c_S$, a spectacular effect due to the
ratio $c_S/v_S<<1$. Roughly speaking,  the Luttinger
fermion has the time to circle many times around the loop
before emitted phonons escape whereas it barely has the
time to move in the electromagnetic case. This ``averaging effect'' explains why the
dependence in the initial relative position is much
weaker for acoustic than for electromagnetic decoherence.
The maximal inverse decoherence rate can be
expressed as an integral in the limit $L>>av_S/c_S$
($k_B\Theta_D=\hbar \omega_D)$:
\begin{equation}
\tau_L\,\ldotp \gamma^{(R/R)}(T) = 
\Delta_{n,m}\,
\frac{g_{ph}^{(d)}(a)}{4^{d-2}\pi}
\,\left(\frac{c_S}{v_S}\right)^2\,\frac{L}{a}
\int_0^1x^{d-1}\,\coth{\left(\frac{x\,\Theta_D}{2T}\right)}\,dx
\end{equation}
The temperature dependence is very weak (remember we are
typically working in situations where $\omega_D\tau_L
>10^5$
and $k_BT\simeq \hbar \omega_L$). To be precise, it goes
like:
\begin{equation}
\frac{\gamma^{(R/R)}(T)-\gamma^{(R/R)}(0)}{\gamma^{(R/R)}(0)}
\simeq \left(\frac{T}{\Theta_D}\right)^{d}
\end{equation}
In opposition with the 
photon bath case, the total acoustic decoherence
time scales as $L^{-1}$ in units of
$\taulutt$. 

\subsection{Caldeira-Leggett computations}
\label{secCaldeira}

The Caldeira-Leggett model corresponds to the Ohmic
spectral density: at low frequencies,
 $\mathcal{J}(\omega) = {M\gamma \omega\over \pi}$. 
In this case, for time scales large
compared to the cut-off time, noise and dissipation
kernels are local in time. The solution of the model is
then much simpler. In the HPZ approach, the equation of
motion defining the $(u_i)_{i=1,2}$ functions can be solved exactly,
taking into
account non perturbatively all effects of
dissipation. For general spectral densities, solutions of
the equation of motions are given in full
generality by a Laplace transform of the form:
$$
\tilde{u}(p)=\frac{\dot{u}(0)+pu(0)}{1+p^2+2\tilde{\eta}(p)}\quad
\mathrm{and}\quad
\tilde{\eta}(p) = 
\int_0^\infty d\omega  \frac{\omega\,\mathcal{J}_l(\omega)}{\omega^2+p^2}.
$$

This expression has clearly a cut along
the $p$-imaginary axis for $|p|< \omega_c$ ($\omega_c$ is
an UV cutoff such that $\mathcal{J}(\omega)=0$ for
$\omega\geq\omega_c$). 
It has no poles in the physical sheet.
The Bromwich contour that encircles
the cut is used to find the inverse Laplace transform~\cite{Alamoudi:1998-1}:
\begin{eqnarray*}
u(t) &=& 2\int_0^{\omega_c}S(\omega) \sin
(\omega t)\, d\omega  \\
S(w) &=&
\frac{\Sigma_I(\omega)}
{(\omega^2-\omega_R^2-\Sigma_R(\omega))^2+\Sigma_I^2(\omega)}
\end{eqnarray*}
The self-energy due to the bath is given by:
\begin{equation}
\Sigma_R(\omega) + i\Sigma_I(\omega) =  2 \mathrm{PP} \int 
\frac{\omega' \mathcal{J}
  (\omega')}{\omega^2-{\omega'}^2}\, d\omega' + i\,
\pi\, \text{sgn}(\omega)\, \mathcal{J}(|\omega|).
\end{equation}
In the weak coupling regime
($\gamma \taulutt <<1$), a Breit-Wigner approximation
can be performed since  $\Sigma_I(\omega_R) <<
\omega_R^2$. Within this approximation, the $u_i$ functions
can be approximated by standard damped oscillator
solutions. In fact,
one could then imagine to use an effective
Caldeira-Leggett model in
order to estimate the decoherence properties of all
Luttinger modes. 

\medskip

In the electromagnetic case,
the decoherence process for the $l>1$ modes takes place
on a typical time scale of order $\tau_l\simeq
\tau_1(c/v_S)^{2l-2}$
which is much longer than for the Ohmic $l=1$ modes. 
In our particular
problem, using asymptotics \ref{eqAsympTotal},
electromagnetic dissipation for the dominant $l=1$ modes is given by
$\gamma= g\tau_L^{-1}/3$. An effective
Caldeira-Leggett model for this mode can then be used. 
In the acoustic case, for a
two dimensional phonon bath and within the relevant range of
frequencies $ \omega_L<\omega<\omega_D$, the spectral
density $\mathcal{J}_l(\omega)$ is ohmic for any $l$, the
Caldeira-Leggett model can be used to describe the
long time dependence of decoherence. 

\medskip

Although the HPZ method can be used to perform explicit
computations, an approximate 
master equation can often be used to obtain the time
evolution of the density matrix of a small subsystem 
$\mathcal{S}$ coupled to a ``reservoir'' $\mathcal{R}$. 
The usual approximation consists first in assuming that
the state of the reservoir is unaltered by the coupling
to the system, and to forget correlations between
$\mathcal{S}$ at time $t'\leq t$ and $\mathcal{R}$ at
time $t$. The second approximation usually done consists
in neglecting non-Markovian terms which can occur for
example in the noise kernel (the dissipation kernel is 
always local in the CL model). 
The condition for these two
approximations to be valid is \cite{Cohen:CDF:1989}:
\begin{equation}
\label{eqRetrMouv}
\tau_c^2 \,\ldotp\,
\mathrm{Tr}\,(H_{\mathcal{SR}}^2\,\rho_{\mathcal{SR}}(0))\,
<< \hbar^2
\end{equation}
where $H_{\mathcal{SR}}$ denotes the coupling between the
system and the reservoir, $\rho_{\mathcal{SR}}(0)$ the
initial total
density operator and $\tau_c$ the correlation time of 
the reservoir.
In the present context, this condition can be expressed as:
\begin{equation}
\label{eqCriterePilot}
\xi_l\,\int_0^{+\infty}\mathcal{J}_l(x/\tau_E)
\coth{\left({l(\beta)\over 2L}\,x\right)}\, dx << 
\left({\tau_E\over \tau_c}\right)^2
\end{equation}
where $\xi_l$ is a dimensionless coefficient that
characterizes the spreading of the $l$-th mode initial
state~:
$\xi_l=1$ for any coherent state and $\xi_l=\coth{(\pi
  \hbar\,lv_S\,\beta /L)}$ for thermal equilibrium.

Within the temperature range used here $T/T_L\simeq 0 -
10$ ($T_L=2\pi \hbar v_S/k_BL$), we have 
$l(\beta) >> L$ for the electromagnetic bath. Therefore, an estimate of the
correlation time for the QEM field is given by the
thermal time. High frequency asymptotics
\ref{eqLimitTE} show that temperature dependence
has negligible effect on the l.h.s. of
\ref{eqCriterePilot} (the effect could be important only
for the $l=1$ mode).
Then, within our temperature range, the l.h.s is approximately bounded by
$gL/a$. The r.h.s is bounded by $L/a$ at high
temperatures and $(v_S/c)\,(T/T_L)$ at lower temperatures.
Therefore, 
since $g\simeq 10^{-8}$, the validity condition is valid
for the photon bath at temperatures above
$g\,(L/a)\,(c/v_S)\times T_L$.
In the acoustic case, the l.h.s. temperature dependance
is only relevant for values of $x$ below
$(c_S/v_S)\times (T/T_L)$. But the main contribution to
the l.h.s comes from higher frequencies is of order
$g_{ph}(L)\,(L/a)^d\simeq g_{ph}(a)\,(L/a)$. At low
temperatures, the r.h.s can be bounded by $(c_S/v_S)\,(T/T_L)$ using
$\hbar \beta c_S$ as an upper value for
$\tau_c$. Condition \ref{eqCriterePilot} can therefore be
rewritten as $(Lv_S/c_Sa)\,g_{ph}(a) <<
T/T_L$. Subsequent computations will assume that the
acoustic UV cutoff is much higher than the typical
Luttinger liquid frequency $2\pi v_S/L$, {\it i.e} that
$Lv_S/c_Sa >>1$. The master equation can therefore only
be used for quite high temperatures.

\medskip

In the case of an underdamped oscillator, as pointed out
by C.~Cohen-Tannoudji \cite{Cohen:CDF:1989}, the approximate
Markovian description can be used provided the noise
kernel is local compared to the observation time
scale. In the present case, this means $\gamma <<\omega_0$ and
$k_BT>>\hbar \gamma$. In the limit of very
low temperatures $k_BT<<\hbar \gamma$, this description
is still valid but one could expect very long time
algebraic tails which precisely take into account non
Markovian effects induced by the divergence of the bath's
correlation time $\hbar\beta$. This possibility is
discussed in appendix \ref{app:matrix} where
these corrections are computed in a weakly damped regime and
shown to be extremely weak.

\subsection{Long time decoherence (Markovian master
  equation treatment)}
\label{secEqPilot}

Within the secular approximation, the master equation can
be translated in the following evolution equation
for the characteristic function. In the case of a single
oscillator of renormalized eigenfrequency $\Omega$ coupled to an Ohmic
bath with dissipation rate $\gamma$:
$Z_t(\lambda,\bar{\lambda})=\mathrm{Tr}(\rho_{\mathcal{S}}(t)\,e^{\lambda
  a^{\dagger}}\,e^{\bar{\lambda}a})$ is given by:
\begin{equation}
\label{eqPilote}
\left[
{\partial \over \partial_t} -
\left(i\Omega-{\gamma\over 2}\right)
\lambda {\partial\over \partial\lambda}
+\left(i\Omega+{\gamma\over 2}\right)
\bar{\lambda}{\partial\over \partial\bar{\lambda}}
\right]
\,Z_t(\lambda,\bar{\lambda})= {\gamma\over
  e^{\beta\hbar\Omega}-1}\,
Z_t(\lambda,\bar\lambda)
\end{equation}
The general solution of this equation is given by:
\begin{equation}
Z_t(\lambda,\bar\lambda) = Z_0\left(\lambda e^{-{\gamma t\over
    2}+i\Omega t},\bar\lambda
e^{-{\gamma t\over 2}-i\Omega t}
\right)\ldotp
\exp{\left(
{1-e^{-\gamma t}\over e^{\beta\hbar\Omega}-1}\,\lambda\bar\lambda
\right)}
\end{equation}
Such a formula immediately shows that, at zero
temperature, coherent states remain coherent but their
parameters evolve according to $\alpha (t) =
\alpha(0)\,e^{-{\gamma t\over 2}-i\Omega t}$. A coherent
superposition of two coherent states decohere as follows:
\begin{eqnarray}
|\alpha\rangle\langle\beta| &\mapsto & |\alpha(t)\rangle
\langle\beta(t)|\times e^{i\theta(t)}\, e^{-d(t)}\\
\theta(t) & = & (1-e^{-\gamma t})\,\Im{(\alpha\bar\beta)}\\
d(t) & = & {|\alpha-\beta|^2\over 2}\, (1-e^{-\gamma t}) 
\end{eqnarray}
The decoherence time is therefore given by a very simple
formula:
\begin{equation}
\tau_{\mathrm{Dec}}^{-1} = \gamma \ldotp
{|\alpha-\beta|^2\over 2}
\end{equation}
Using formula \ref{eqDecohCoeff} in the Caldeira-Leggett case provides the same
result and the temperature dependence is given, in the
limit $\gamma/2\Omega_1\rightarrow 0$ ($\Omega_1=\sqrt{\Omega^2-\gamma^2/4}$) by:
\begin{equation}
D_T(t) = \frac{\Theta(T)\, (1-e^{-\gamma\,t})}
{1+(\Theta(T)-1) \ldotp(1-e^{-\gamma\,t})}\times
\boldsymbol{1}\quad
\mathrm{with}\quad 
\Theta (T)
=\coth{\left(\frac{\beta\hbar\Omega_1}{2}\right)}
\end{equation}
Therefore, the $t\rightarrow +\infty$ limit of
decoherence is independent of temperature but the
decoherence time will scale with temperature according to
the $\Theta(T)$ factor: 
\begin{equation}
\label{eqDecohTemp}
\frac{\tau_{Dec}(T)}{\tau_{Dec}(0)}\simeq
\tanh{\left(\frac{\hbar\Omega_1 }{2k_BT}\right)}
\end{equation}

\subsection{Decoherence at long time}
\label{secLongTime}

Applying previous results to the {\em electromagnetic
decoherence} ($l=1$ modes), we obtain:
\begin{eqnarray}
d_{\infty}^{(R/R)}(\sigma_1,\sigma_2) & = & 
4\Delta_{n,m}\,\ldotp\,
\sin^2{\left({\pi \sigma_{12}\over L}\right)} \\
d_{\infty}^{(L/R)}(\sigma_1,\sigma_2) & = & 
2\Delta_{n,m}\,\ldotp\,\left(
1+{m^2-4n^2\alpha^2\over
  m^2+4n^2\alpha^2}
\cos{\left({2\pi\sigma_{12}\over L}\right)}\right)
\end{eqnarray}
As expected, the $R/R$ limiting decoherence vanishes for
$\sigma_1=\sigma_2$ whereas the $R/L$ one does not. The
typical asymptotic value is proportional to 
$\Delta_{n,m}$. In fact, this number can be viewed as
measuring the ``distance'' between the two
quantum states which built our Schr\H{o}dinger cat. Using
vertex operator with small values of$m$ and $n$ in Schr\H{o}dinger cats
\ref{defCatRR} and \ref{defCatRL} produces mesoscopically separated
coherent states in each mode.

\medskip

Figure 6 summarizes the 
electromagnetic decoherence
exponent as a function of time for $\gamma\,t\simeq 0 -
5$ and for various temperatures.


In the {\em acoustic case (two dimensional phonon bath)}, contributions
of all relevant modes should be summed. As before, the
$\sigma_{12}$ dependence disappears as soon as
$\sigma_{12}>>av_S/c_S$. 
Introducing $T=2\pi^2\tau_Lv_S/(c_Sg_{ac}(a))$, the sum
over all Luttinger modes up to the cut-off frequency can be evaluated:
\begin{equation}
\frac{d^{(R/R)}(t)}{d^{(R/R)}(\infty)} = 
\frac{d^{(R/L)}(t)}{d^{(R/L)}(\infty)} = 
1+\frac{e^{-t/T}-1}{(t/T)}
\end{equation}
where 
$d^{(R/R)}(\infty)=d^{(R/L)}(\infty)=\Delta_{n,m}\,c_SL/(v_Sa)$. For the
continuum approximation to be valid, we have assumed that
$L/a>>v_S/c_S$ and therefore
$d^{(R/R)}(\infty)>>1$. This also implies that most of
the decoherence process is accomplished within the
previously computed acoustic decoherence time
$2T/d^{(R/R)}(\infty)<<T$.

\section{Conclusion and discussion}
\label{secConclu}

Within the bosonization framework, 
we have shown how the coupling to an
external quantum electromagnetic field or to a two or
three dimensional bath of longitudinal phonons can lead to
decoherence of a Schr\H{o}dinger cat state formed of
localized elementary excitations in a Luttinger liquid. 
Three different phases build  the decoherence scenario for
Schr\H{o}dinger cat states of the Luttinger liquid (see
7 for the electromagnetic case):

\begin{itemize}

\item At a very short time, because of the time/energy
  uncertainty relation, energy exchanges between
 the environment and the Luttinger system are not conservative. In
 this regime, high frequencies take part in the
 decoherence process and non-Markovian effects are
 important. The precise time evolution of the system is
 strongly cut-off dependent.

\item After a transitory regime, the decoherence
  exponent becomes linear in time. This is the ``Golden
  Rule'' regime: energy conservation between the
  environment field and the Luttinger system is
  satisfied with a spectral width going down to its
  natural value (spontaneous photon or phonon emission). Only
  frequencies in resonance with the Luttinger
  eigenfrequencies contribute to dissipation. 

The structure of spectral densities for the
environmental modes leads to a hierarchy of
decoherence times corresponding to the multipolar
expansion of the ``radiation'' emitted by the Luttinger
system. In the acoustic case, the various decoherence
times do not increase as much with increasing $l$. In the
electromagnetic case, they increase with $l$. The
total decoherence time is therefore much smaller in the
acoustic than in the electromagnetic case.

\item At longer times, decoherence tends to saturation.
In this regime, an effective Caldeira-Leggett
model can be used to describe the dominating
decoherence processes: in the electromagnetic case, one
can keep only $l=1$ modes, corresponding 
to dipolar electromagnetic radiation. In the 2D
acoustic case, one can use an effective Caldeira-Leggett
model for all modes.
Caldeira-Leggett computations are non-perturbative
since they take into account all orders of the coupling
between the quantum environment and the oscillator. 

\medskip

The infinite time decoherence depends on two factors. The first
one is, as expected, the distance between the two quantum states
which depends on the dimension of the operators used in
these states. The 
second one reflects the relative weight of each
hydrodynamic mode in the decoherence process. In the
electromagnetic case, since $l=1$ modes dominate, it
gives a geometrical factor depending on the relative
position $\sigma_{12}$. In the acoustic case, spatial
dependence is almost always lost and we are left with an
important mode number factor. That's why, although
$\psi^{\dagger}_R(\sigma_1)|0\rangle$ and
$\psi^{\dagger}_R(\sigma_1)|0\rangle$ can be considered
as ``mesoscopically close'' with respect to 
their ``distance'', decoherence is much faster than
dissipation in the acoustic case. This is a major
difference with single mode decoherence studies
\cite{CL:1985-1} where the ratio between decoherence and
dissipation times is only due the distance between states
entering the Schr\H{o}dinger cat.

\end{itemize}



Non-linearities in the spectrum of low
energy excitations may also contribute to decoherence. As
showed by Haldane \cite{Haldane:1981-1},
non-linearities in the spectrum couple the bosonic modes of the theory,
turning the simple free model used in bosonization into
an interacting theory. Coupling between modes should
also play an important role in the decoherence properties of
Schr\H{o}dinger cat states. Indeed, the model presented
here provides an upper limit for decoherence times.
Non-linearities should be also taken into
account when investigating the resistive behavior
of small metallic loops induced by inelastic collisions 
\cite{Landauer:1985-1}. 

\medskip

In the present work, photons or longitudinal phonons
initially
at equilibrium were used as the thermal bath for the
system, but of course, one could imagine various
extensions. One could use another description for quantum fluctuations 
of the environment. For example, one may think about
changing the state of the environment, taking for example into account an
external microwave radiation. Increasing the incoming
radiation power within the range of resonant frequencies should
increase decoherence of Schr\H{o}dinger cat states
(enhancement of dissipation by stimulated emission of radiation).
With such environmental states, one expects to meet
also the problem of ``decoherence repartition'' between all the modes of the
Luttinger system (even in the electromagnetic case). 
Although this makes computations much
harder to control, it may lead to more
interesting behaviors. Finally, we are also investigating
the extension of these ideas to two or three dimensional systems.

\begin{acknowledgments}
We acknowledge many useful discussions and encouragements
with Ch.~Chaubet, L. Cugliandolo, B. Dou\c{c}ot,
J.~Kurchan  and
L.~Saminadayar. A.~De~Martino, R.~M\'elin and P.~Pujol are
warmly thanked for their careful reading of the
manuscript and their numerous remarks and questions. 
D.~Mouhanna has kindly sent us a copy of
C.~Cohen-Tannoudji lectures on decoherence at Coll\`ege de France.

Part of this work was performed during the Spring
workshop on {\em Strongly correlated fermions} at the
Institut Henri Poincar\'e (Apr. - July 1999). The
organizers of the XXIVth Rencontres de Moriond on {\em
Quantum Physics at mesoscopic scale} are thanked for giving us the
opportunity to present this work. We also thank D.~Loss
and Th.~Martin for stimulating conversations during this meeting and
communication of many useful references.
\end{acknowledgments}

\appendix
\section{Keldysh's Green's functions}
\label{secAppGreen}

Let us recall the well known expressions for a
single harmonic mode of frequency $\omega$ at temperature
$\beta$ (unit mass) \cite{Mahan}:

\begin{eqnarray}
G_>(t,s) & = &
\frac{-i}{2\omega}\left(\coth{(\frac{\beta\hbar\omega}{2})}
\cos{(\omega(t-s))}-i\sin{(\omega(t-s))}\right)\\
G_<(t,s) & = &
\frac{-i}{2\omega}\left(\coth{(\frac{\beta\hbar\omega}{2})}
\cos{(\omega(t-s))}+i\sin{(\omega(t-s))}\right)\\
G_T(t,s) & = &
\frac{-i}{2\omega}\left(\coth{(\frac{\beta\hbar\omega}{2})}
\cos{(\omega(t-s))}-i\sin{(\omega\,|t-s|)}\right)\\
G_{\tilde{T}}(t,s) & = &
\frac{-i}{2\omega}\left(\coth{(\frac{\beta\hbar\omega}{2})}
\cos{(\omega(t-s))}+i\sin{(\omega\,|t-s|)}\right)\\
\end{eqnarray}
The retarded and advanced Green functions can be related
to these expressions by
\begin{equation}
G_R = G_T-G_<\quad\mathrm{and}\ G_A=G_T-G_>
\end{equation}
For the electromagnetic field, a mode decomposition can
be used. Using results of appendix \ref{secAppModes}, we 
obtain:
\begin{equation}
\boldsymbol{D}^{\alpha\beta}((\boldsymbol{x},t);(\boldsymbol{y},s))=
\sum_I\boldsymbol{A}_I^{\alpha}(\boldsymbol{x})
\boldsymbol{A}_I^{\beta}(\boldsymbol{y})
\, \boldsymbol{G}_{\omega_I}(t,s)
\end{equation}
The noise and dissipation kernels \ref{eqNoise} and \ref{eqDissip} can be
expressed as:
\begin{eqnarray}
\label{eqNoiseExpl}
\nu^{\alpha\beta}(x,y) & = & \sum_I\boldsymbol{A}_I^{\alpha}(\boldsymbol{x})
\boldsymbol{A}_I^{\beta}(\boldsymbol{y})
\,\coth{\left(\frac{\beta\hbar\omega}{2}\right)}
\frac{\cos{(\omega_I(t-s))}}{2\omega_I}
\\
\label{eqDissipExpl}
\eta^{\alpha\beta}(x,y) & = & 
\sum_I\boldsymbol{A}_I^{\alpha}(\boldsymbol{x})
\boldsymbol{A}_I^{\beta}(\boldsymbol{y})\,
\frac{\sin{(\omega_I(t-s))}}{2\omega_I}
\end{eqnarray}

\section{Electromagnetic modes in a cylindrical cavity}
\label{secAppModes}

We shall decompose the transverse vector potential in
orthonormal modes:
\begin{equation}
\boldsymbol{A}(\boldsymbol{r},t)=\sum _I\psi_I(t)
\boldsymbol{A}_I(\boldsymbol{r})
\end{equation}
These modes can be real, or complex. In the latter case,
we assume the existence of an involution $I\mapsto
\hat{I}$ over
the set of indices implementing complex conjugation: 
$(\boldsymbol{A}_I(\boldsymbol{r}))^*=\boldsymbol{A}_{\hat{I}}(\boldsymbol{r})$.
Modes are
normalized with respect of the volume ${\cal V}$ of the cavity by
imposing an orthogonality condition:
$$\int d^3\boldsymbol{r}\;
\boldsymbol{A}_I(\boldsymbol{r})\ldotp
\boldsymbol{A}_J(\boldsymbol{r})
=\delta_{I,\hat{J}}.$$
The $\psi_I(t)$ are coordinates for a set of 
{\em independent} harmonic oscillators of frequency $\omega_I$.

\medskip

Expressions are given in cylindrical coordinates
$(r,\vartheta,z)$.
$R$ denotes the radius of the cavity. 
It is useful to introduce:
$\boldsymbol{A}(\boldsymbol{r})=
\boldsymbol{{\cal A}}(\boldsymbol{r})/\sqrt{{\cal V}}$.

\begin{description}

\item[TE modes] Let $l$ be the angular momentum of the
  mode around the $Oz$ axis and $k_k$ its momentum along
  this axis. Let  $u'_{l,n}$ denote the $n$-th zero
  of the derivative of the $l$-th order Bessel function
  $J_l$. 
Then, the orthoradial component of the TE $(l,k_z,n)$
complex mode is given by:

\begin{equation}
\label{eqModeTE}
\boldsymbol{{\cal A}}_{\vartheta} = 
-\frac{u'_{l,n}}{\sqrt{(u'_{l,n})^2-l^2}}
\frac{J'_l(\frac{r}{R}u'_{l,n})}{J_l(u'_{l,n})}\,
e^{i(l\vartheta-k_zz)}
\end{equation}
We have $\omega=c\sqrt{k_z^2+(u'_{l,n}/R)^2}$.

\item[TM modes] Here  $u_{l,n}$ denotes the $n$-th zero
  of the $l$-th order Bessel function
  $J_l$. 
Then, the orthoradial component is given by:

\begin{equation}
\label{eqModeTM}
\boldsymbol{{\cal A}}_{\vartheta} =
-\frac{k_zR}{
\sqrt{u_{l,n}^2+(k_zR)^2}}
\frac{l}{u_{l,n}}
\frac{J_l(\frac{r}{R}u_{l,n})}{J'_l(u_{l,n})}\,
e^{i(l\vartheta - k_zz)}
\end{equation}
We have $\omega=c\sqrt{k_z^2+(u_{l,n}/R)^2}$.

\end{description}

Normalized real modes are inferred from these complex modes by:
\begin{equation}
\boldsymbol{{\cal A}}_I^{(+)}= \frac{1}{\sqrt{2}}
\left(\boldsymbol{{\cal A}}_I+\boldsymbol{{\cal A}}_{\hat{I}}\right)
\quad\mathrm{and}\ 
\boldsymbol{{\cal A}}_I^{(-)}= \frac{i}{\sqrt{2}}
\left(\boldsymbol{{\cal A}}_I-\boldsymbol{{\cal
      A}}_{\hat{I}}\right)
\end{equation}
Expressions \ref{eqModeTE} and \ref{eqModeTM} show that:
\begin{equation}
  \label{eqRealModes}
\begin{cases}
  \boldsymbol{{\cal
      A}}_{I,\vartheta}^{(+)}=\mathcal{D}_I\cos{(l\vartheta-k_zz)}\\
\boldsymbol{{\cal
    A}}_{I,\vartheta}^{(-)}=-\mathcal{D}_I\sin{(l\vartheta-k_zz)}
\end{cases}
\end{equation}
where $mathcal{D}_I$ contains the $r$ dependence and depends on
the mode characteristics $TE/TM$ and $(l,k_z,n)$. More
precisely we have
\begin{eqnarray}
\label{eqCoeffDTE}
TE\ \mathrm{modes:} \quad \mathcal{D}_I & = & 
-\frac{u'_{l,n}\sqrt{2}}{\sqrt{(u'_{l,n})^2-l^2}}\,
\frac{J'_l(\frac{r}{R}u'_{l,n})}{J_l(u'_{l,n})}
\\
\label{eqCoeffDTM}
TM\ \mathrm{modes:}\quad \mathcal{D}_I & = & 
\frac{Rk_z\sqrt{2}}{\sqrt{u_{l,n}^2+(k_zR)^2}}\,
\frac{l}{u_{l,n}}\,
\frac{J_l(\frac{r}{R}u_{l,n})}{J'_l(u_{l,n})}
\end{eqnarray}

\section{Longitudinal acoustic modes in a cylindrical cavity}
\label{secAppModesAcoustic}

The displacement field $\boldsymbol{u}(\boldsymbol{x},t)$
contains a gradient part:
$$\boldsymbol{u}(\boldsymbol{x},t)= \sum_I
\frac{Q_I(t)}{\sqrt{\mathcal{V}}}\,
(\boldsymbol{\nabla}\varphi_I)(\boldsymbol{x})
$$
where the $\varphi_I$ functions are eigenvalues of the
Laplacian with von~Neuman's boundary conditions. The
modes $\boldsymbol{u}_I=\boldsymbol{\nabla}\varphi_I$ are
normalized so that
$$\int d^3\boldsymbol{r}\;
\boldsymbol{u}_I(\boldsymbol{r})\ldotp
\boldsymbol{u}_J(\boldsymbol{r})
=\delta_{I,\hat{J}}\,\mathcal{V}.$$
The $Q_I(t)$ are coordinates for a set of 
{\em independent} harmonic oscillators with given
frequency $\omega_I$.
Longitudinal acoustic modes are indexed by $I=(l,k_z,n)$
for $d=3$ and $I=(l,n)$ for $d=2$ and are typically of the form
$e^{i(l\theta-k_zz)}\mathcal{N}_I$ where 
$\mathcal{N}_I$ contains the $r$ dependence 
($\omega_I=c_S\,\sqrt{k_z^2+(u'_{l,n}/R)^2}$ for $d=3$
and $\omega_I=c_Su'_{l,n}/R$ for $d=2$): 
\begin{eqnarray}
\label{eqNCoeff2}
d=2: \quad \mathcal{N}_I & = & 
\frac{2\,u'_{l,n}}{(\omega_I\tau_S)\,\sqrt{(u'_{l,n})^2-l^2}}\,
\frac{J_l(\frac{r}{R}u'_{l,n})}{J_l(u'_{l,n})}
\\
\label{eqNCoeff3}
d=3:\quad \mathcal{N}_I & = & 
\frac{u'_{l,n}\sqrt{2}}{(\omega_I\tau_S)\,\sqrt{(u'_{l,n})^2--l^2}}\,
\frac{J_l(\frac{r}{R}u'_{l,n})}{J_l(u'_{l,n})}
\end{eqnarray}

\section{Asymptotic behavior of spectral densities}
\label{secAsymptotic}

Asymptotic expressions of the spectral densities corresponding to the
infinite-cavity limit are useful to study the long time 
behavior of the decoherence and dissipation processes. 
The radial quantization is of $\Delta \omega_\perp =
\Delta u_{l.n}^{(')} c/R \simeq \pi c/R$ and the
longitudinal one $\Delta \omega_\parallel = 2 \pi c /h$.
We shall therefore use the following expressions for the extremas of the
Bessel functions:
\begin{equation}
J_l(u_{l,n}')\simeq \sqrt{\frac{2}{\pi u_{l,n}'}} \qquad
\text{and} \qquad J_l '(u_{l,n})\simeq \sqrt{\frac{2}{\pi
    u_{l,n}}}
\end{equation}
and the following low- and high- frequency expansions
\begin{equation}
J_l(z) \simeq \frac{ z^l}{2^l l!} \quad\mathrm{and}\ 
J_l(z) \simeq  \sqrt{\frac{2}{\pi z}} \cos(z-l\pi /2-\pi /4)
\end{equation}
A straightforward algebra gives us:
\begin{enumerate}
\item In the {\em low-frequency regime} ($\omega\,\taulight <<1$)
\begin{align}
\label{eqAsympTE}
\mathcal{J}_{l \neq 0}^{TE}(\omega) &\simeq \frac{g}{\pi}\;
\;\frac{l}{(2l-1)!} \;\Big(
\frac{\omega \taulight}{2\pi} \Big)^{2l-1} \\
\label{eqAsympTEZero}
\mathcal{J}_0^{TE}(\omega) &\simeq \frac{1}{3\pi}\;
\frac{v_S}{c}\; \frac{\alpha_{QED}}{\alpha}\; \Big(
\frac{\omega \taulight}{2\pi} \Big)^3 \\
\label{eqAsympTM}
\mathcal{J}_{l \neq 0}^{TM}(\omega) &\simeq \frac{g}{\pi}\; 
\frac{2l^2}{(2l+1)!}\;\Big(
\frac{\omega \taulight}{2\pi} \Big)^{2l-1} \\
\label{eqAsympTMZero}
\mathcal{J}_0^{TM}(\omega) &=0
\end{align}
Henceforth, the total spectral densities behave as:
\begin{align}
\label{eqAsympTotal}
\mathcal{J}_{l \neq 0}(\omega) &\simeq \frac{g}{\pi}\;
\frac{l^2(l+1)}{(2l+1)!} \;\Big(
\frac{\omega \taulight}{2\pi} \Big)^{2l-1} \\
\label{eqAsympZero}
\mathcal{J}_0(\omega) &\simeq \frac{1}{3\pi}\;
\frac{v_S}{c}\; \frac{\alpha_{QED}}{\alpha} \;\Big(
\frac{\omega \taulight}{2\pi} \Big)^3
\end{align}
\item In the {\em high-frequency regime} ($\omega\,\taulight >>1$)
\begin{align}
\label{eqLimitTE}
\mathcal{J}_l^{TE}(\omega) &\simeq
{l\,g\over 2\pi} \\
\label{eqLimitTM}
\mathcal{J}_l^{TM}(\omega) &\simeq   \omega^{-1}
\end{align}
\end{enumerate}

These analytic results agree with the numerics, which are
represented in figure 3.
The main result is that all the modes are supraohmic at
low frequency,
expect the mode $l=1$ which shows ohmic behavior.

\medskip

The same kind of expansion can be performed for the
acoustic spectral densities. In this case, one obtains
($d\in \{2,3\}$):

\begin{itemize}

\item In the {\em low frequency regime}:

\begin{equation}
\mathcal{J}_l(\omega) \propto g_{Ph}(L)\,
(\omega\tau_S)^{2l+d}
\end{equation}

\item In the {\em high frequency regime}:

\begin{equation}
\mathcal{J}_l(\omega) \simeq g_{Ph}(L)\,\frac{l}{2^{2(d-2)}\pi}\,
(\omega\tau_S)^{d-1}
\end{equation}

\end{itemize}

Only the latter will be used to compute decoherence
properties since $\omega_l$ falls into the high frequency
regime since $v_S>>c_S$. 

\section{Decoherence matrix computations for a
  Caldeira-Leggett solution}
\label{app:matrix}

The appendix presents details of the computation of the decoherence
matrix $D$ (\ref{eqDecohCoeff}) in the case of a damped
oscillator solution. This has direct relevance for the
Caldeira-Leggett model but also within the framework of 
the Breit-Wigner approximation in more general environments.
These results have been discussed in
\cite{CL:1985-1,Hu:1995-1} but their derivation is
recalled here in a simpler way. Strictly speaking the
approximate master equation approach described in section
\ref{secCaldeira} is not valid for temperatures below the
$\hbar\gamma /k_B$. Therefore, 
this appendix aims at finding ultra-low temperature
corrections to decoherence arising from the non-Markovian
effects arising from the $T\rightarrow 0$ behavior of 
the reservoir symmetric two point correlation function.

\medskip

Let $\Omega_R$ and $\gamma$ denote the renormalized frequency of
the oscillator and $\gamma$ the damping
coefficient. We shall work in the weakly damped case,
defined by $\Omega_1^2=\Omega_R^2-\gamma^2/4\geq 0$, and
measure the strength of dissipation by $\phi$ such that
$\tan{(\phi)} =\gamma/2\Omega_R$. The decoherence matrix
can then be expressed in terms of the following auxiliary
functions:
\begin{eqnarray}
Z_{\pm}(t,\omega) &=& \int_0^t e^{-\gamma s/2 +i(\omega \pm
  \Omega_1)s} ds, \\
S(t) &=& \int _0^\infty (|Z_+|^2 +
|Z_-|^2)(t,\omega) \mathcal{J} (\omega) \coth{\left(\frac{\beta
\hbar \omega}{2}\right)}\,d\omega , \\
P(t) &=& \int _0^\infty (Z_+
\overline{Z_-})(t,\omega) \mathcal{J} (\omega) \coth{\left(\frac{\beta
\hbar \omega}{2}\right)}\,d\omega ,
\end{eqnarray}
We have:
\begin{eqnarray}
D &=& \frac{1}{d} \left( \begin{array}{cc}
D_{11} & D_{12} \\
D_{12} & D_{22} \end{array} \right),\\
d &=& 4 \left( 1+ \frac{2 e^{-\gamma t}}{\Omega_R(S^2 -4P\bar{P})}
  \left( S\left(\frac{\Omega_R}{\Omega} +
      \frac{\Omega}{\Omega_R}\right)+2\,\Re \left( P e^{-2i\Omega_1 t}
      \left( \frac{\Omega_R}{\Omega} e^{-2i\phi} -
        \frac{\Omega}{\Omega_R} \right) \right) \right)
 \right. +\\
&+&\left. \frac{16 \cos^2{(\phi)}\,e^{-2\gamma t}}{\Omega_R^2 
      (S^2 - 4 P\bar{P})} \right),
\end{eqnarray}
and the coefficients are given by:
\begin{eqnarray}
D_{11} &=& \Omega + \frac{2 e^{-\gamma t}}{S^2
  -4 P\bar{P}} (S+2\,\Re(P e^{-2i(\Omega_1 t + \phi)})) ,\\
D_{12} = D_{21} &=& \frac{2 e^{-\gamma t}}{ \Omega_R(S^2
  -4 P\bar{P})} (S \sin{(\phi)} + 2\,\Im (P e^{-i(2\Omega_1 t
  +\phi)})) ,\\
D_{22} &=& \frac{1}{\Omega} + \frac{2  e^{-\gamma t}}{ \Omega_R(S^2
  -4 P\bar{P})} (S+2\,\Re(P e^{-2i\Omega_1 t})).
\end{eqnarray}

The $S$ and $P$ functions can be computed by the residue theorem:
\begin{eqnarray}
S &=& \frac{1}{2} \int_{-\infty}^\infty\left(
  \frac{1+e^{-\gamma t} -2e^{-\gamma t/2} e^{i (\omega
      +\Omega_1)t}}{(\omega+\Omega_1)^2+\gamma^2 /4} + \frac{1+e^{-\gamma t} -2e^{-\gamma t/2} e^{i (\omega
      -\Omega_1)t}}{(\omega-\Omega_1)^2+\gamma^2 /4}
\right) \mathcal{J}(\omega) 
\coth{\left(\frac{\beta \hbar \omega}{2}\right)}\, d\omega 
,\\
P &=& \frac{1}{2} \int_{-\infty}^\infty 
\frac{1+e^{-\gamma t} e^{2i\Omega_1 t} - 2 e^{-\gamma
    t/2} e^{i\Omega_1 t} e^{i\omega
    t}}{(\omega+\Omega_1+i\gamma /2)(\omega-\Omega_1 -i\gamma/2)}
\mathcal{J}(\omega) \coth{\left(\frac{\beta \hbar \omega }{2}\right)}\,d\omega.
\end{eqnarray}

The main contribution to $S$ and $P$ is due to
the poles $\pm \Omega_1 +i\gamma /2$:
\begin{eqnarray}
\label{eqSDiss}
S^{(Main)} &=& \frac{2\pi}{\gamma} (1-e^{-\gamma t}) \,
\Re{\left(\mathcal{J}(\Omega_1 +i\gamma/2) \coth{\left(\frac{\beta \hbar}{2} (\Omega_1 +i\gamma
/2)\right)}\right)} ,\\
\label{eqPDiss}
P^{(Main)} &=& \frac{i \pi}{2} (1-e^{-\gamma t} e^{2i\Omega_1 t})\,
\frac{\mathcal{J}(\Omega_1 +i\gamma/2)}{\Omega_1 +i\gamma
  /2}\, \coth{\left(\frac{\beta \hbar}{2} (\Omega_1 +i\frac{\gamma}{2})\right)},
\end{eqnarray}
In this expression cut-off dependent quantities have been
discarded since they can be shown to be of order
$(\gamma/2\Omega_1)\log{(\Lambda/\Omega_R)}$, {\it i.e.} much smaller than
the oscillator eigenfrequency's renormalization.
The temperature dependent part also contain poles which
give a series of exponentially decreasing terms of the
form $\exp{(-2\pi nt/\tau_{Th})}$ ($n\geq 1$). 
At vanishing temperature, this series can be resummed
into an algebraically decreasing correction as follows
(here $\mathcal{J} (\omega) = \gamma\omega/\pi\Omega_1^2$):
\begin{eqnarray}
\label{eqSAsymp}
S^{(Corr.)} &=& -\frac{2e^{-\gamma t/2}}{\pi \Omega_1^2 t}\,
\Re{\left( e^{i\Omega_1 t} (z_t \mathcal{S} (z_t) +\bar{z_t}
  \mathcal{S} (-\bar{z_t}) ) + e^{-i\Omega_1 t} (z \mathcal{S} (-z_t) +\bar{z_t}
  \mathcal{S} (\bar{z_t}) ) \right)} ,\\
\label{eqPAsymp}
P^{(Corr.)} &=& -\frac{\phi e^{-\gamma t/2} e^{i\Omega_1
    t}}{\pi \Omega_1} \left( \mathcal{S} (z_t) +
  \mathcal{S} (-z_t) - \overline{\mathcal{S} (\bar{z_t})} -
  \overline{\mathcal{S} (-\bar{z_t})}  \right)
\end{eqnarray}
where $\mathcal{S} (z) = e^z\,\mathrm{Ei}(1,z)$
and $z=(\gamma /2 -i \Omega_1)\, t$.
$\mathcal{S} (z_t)$ has an asymptotic expansion in $1/t$ for
$t\rightarrow +\infty$. 

\medskip

At zero temperature, these corrective terms \ref{eqSAsymp} and \ref{eqPAsymp} 
dominate \ref{eqSDiss} and \ref{eqPDiss} for
$\gamma t >>\log(2\Omega_1 / \gamma)$.
The very long time asymptotic of $S$ is therefore 
given by:
\begin{eqnarray}
S \simeq \frac{2}{\Omega_1}\left(1+
\,\frac{4\phi\,e^{-\gamma t/2} \cos{(\Omega_1 t)}}{\pi \,(\Omega_1  t)^2}\right)
\end{eqnarray}
In case of weak damping $\gamma << \Omega_R$, the
following approximations can be made: $\phi <<
1$, $\Omega \simeq \Omega_R \simeq \Omega_1$ and 
$f(\Omega_1 +i\gamma/2) \simeq f(\Omega_1)$.
We will neglect $P$ since, in full generality, $P \simeq
\phi\,S$. Plugging everything in $D$'s expression, one
finally ends up with a very long time asymptotics:
\begin{equation}
D(T=0) \simeq  \left(1-e^{-\gamma
    t} +\frac{4\phi}{\pi}\,e^{-3\gamma t/2}
\,\frac{\cos{(\Omega_1 t)}}{(\Omega_1 t)^2}\right)\,\ldotp \left(
\begin{array}{cc}
\Omega_1/4 & 0 \\ 0& 1/4\Omega_1 
\end{array} 
\right).
\end{equation}
Henceforth, in the very weak damping situation, master
equation results for the decoherence matrix can safely be
extrapolated down to $T=0$, even if strictly speaking
non-Markovian effects should be taken into account.

\bibliography{/home/degio/TeX/Decoherence/biblio,/home/degio/TeX/biblio/Lie} 

\end{document}